\documentclass[graybox]{svmult}

\usepackage{mathptmx}       
\usepackage{helvet}         
\usepackage{courier}        
\usepackage{makeidx}         
\usepackage[pdftex]{graphicx}
\usepackage{multicol}        
\usepackage[bottom]{footmisc}

\usepackage{amssymb}
\usepackage{amsmath,amsfonts,latexsym}
\usepackage{array,tabularx,color}
\usepackage[normalem]{ulem}
\usepackage{comment}
\DeclareMathOperator{\Ry}{\mathcal{R}y}
\newcommand{\vect}[1]{{\mathbf #1}}
\newcommand{\Frac}[2]{\displaystyle\frac{#1}{#2}}

\makeindex             

\begin{document}

\title*{Vortices in polariton OPO superfluids}

\author{Francesca Maria Marchetti and Marzena H. Szyma\'nska}

\institute{
  Francesca Maria Marchetti \at Departamento de F\'isica Te\'orica de
  la Materia Condensada, Universidad Aut\'onoma de Madrid, Madrid
  28049, Spain \email{francesca.marchetti@uam.es}
\and
  Marzena H. Szyma\'nska \at Department of Physics, University of
  Warwick, Coventry, CV4 7AL, UK \email{M.H.Szymanska@warwick.ac.uk}}
%
%
%
\maketitle


\abstract{This chapter reviews the occurrence of quantised vortices in
  polariton fluids, primarily when polaritons are driven in the
  optical parametric oscillator (OPO) regime. We first review the OPO
  physics, together with both its analytical and numerical modelling,
  the latter being necessary for the description of finite size
  systems. Pattern formation is typical in systems driven away from
  equilibrium. Similarly, we find that uniform OPO solutions can be
  unstable to the spontaneous formation of quantised
  vortices. However, metastable vortices can only be injected
  externally into an otherwise stable symmetric state, and their
  persistence is due to the OPO superfluid properties. We discuss how
  the currents charactering an OPO play a crucial role in the
  occurrence and dynamics of both metastable and spontaneous
  vortices.}


\section{Introduction}
\label{sec:intro}
Quantised vortices are topological defects occurring in
macroscopically coherent systems, and as such have been broadly
studied in several area of physics. Their existence was first
predicted in superfluids~\cite{onsager_49,feynman_55}, and later in
coherent waves~\cite{nye74}. Nowadays, quantised vortices have been
the subject of extensive research across several areas of physics and
have been observed in type-II superconductors, ${}^4$He, ultracold
atomic gases, non-linear optical media (for a review see,
e.g.,~\cite{staliunas}) and very recently microcavity
polaritons~\cite{lagoudakis08,lagoudakis09a,lagoudakis10,sanvitto10,krizhanovskii10,roumpos10,nardin11,sanvitto11},
the coherent strong mixing of a quantum well exciton with a cavity
photon.

This chapter reviews the occurrence of quantised vortices in polariton
fluids, primarily when polaritons are driven in the optical parametric
oscillator (OPO) regime. The interest in this area of research is
manifold. To start with, the search for condensation in solid state
excitonic systems has been arduous and lasted more than two decades:
Unambiguous evidence for condensation has been reported for
microcavity polaritons for the first time in
2006~\cite{kasprzak06:nature}. These results have been followed by a
wealth of experimental and theoretical advances on aspects related to
macroscopic coherence, condensation, superfluidity, quantum
hydrodynamics, pattern formation, just to mention few (for a review,
see Ref.~\cite{deng10}).
Two different schemes of injecting polaritons and spontaneously
generating a macroscopically coherent state can be employed: (i)
non-resonant pumping, and (ii) parametric drive in the
optical-parametric-oscillator (OPO) regime. What both condensates have
in common is the phenomenon of spontaneous phase symmetry breaking
(and the consequent appearance of a Goldstone mode), and the
non-equilibrium ingredient. However, the way polaritons are pumped has
strong effects on the type of condensed regime that can be reached. In
both regimes (i) and (ii), the quest for superfluid behaviour has been
and is being widely investigated. How it has been recently discussed
in Refs.~\cite{carusotto_pr,keeling09,keeling_rev11}, one of the
aspects that makes condensed polariton systems novel compared to known
superfluids at thermal equilibrium, is that now all the paradigmatic
definitions of a superfluid, such as the appearance of quantised
vortices, the Landau criterion, the existence of metastable persistent
flow, the occurrence of solitary waves, have to be singularly examined
and might in general be fundamentally different from the equilibrium
case. Several of these popular topics are examined in other chapters
of this book {\tt cross-refer to : A. Bramati and A. Amo, B. Deveaud,
  D. Krizanovskii and M. Skolnick, F. Laussy, G. Malpuech, M. Wouters
  and V. Savona, D. Snoke, Y. Yamamoto}.

Resonantly pumped polaritons in the OPO
regime~\cite{stevenson00,baumberg00:prb} have been recently shown to
exhibit a new form of non-equilibrium
superfluidity~\cite{amo09,sanvitto10}.
Polaritons continuously injected into the \emph{pump} state, undergo
coherent stimulated scattering into the \emph{signal} and \emph{idler}
states.
Superfluidity has been tested through as frictionless
flow~\cite{amo09} of a travelling signal triggered by an additional
pulsed probe laser (the TOPO regime).
In addition, the study of quantised vortices imprinted using pulsed
Laguerre-Gauss laser fields has attracted noticeable interest both
experimentally~\cite{sanvitto10} and
theoretically~\cite{wouters10,marchetti10,szymanska10,gorbach10},
providing a diagnostic for superfluid properties of such a
non-equilibrium system.
In particular, vorticity has been shown to persist not only in absence
of the rotating drive, but also longer than the gain induced by the
probe, and therefore to be transferred to the OPO signal,
demonstrating metastability of quantised vortices and persistence of
currents~\cite{sanvitto10,marchetti10}.


The chapter is arranged as
follows: after a very short introduction to microcavity polaritons in
Sec.~\ref{sec:intpo}, we describe the optical parametric oscillator
regime in Sec.~\ref{sec:opore}, stressing the analogies and
differences with an equilibrium weakly interacting Bose-Einstein
condensate (Sec.~\ref{sec:golds}) and the numerical modelling that is
necessary to use for finite size pumps (Sec.~\ref{sec:numer}). In
Sec.~\ref{sec:stabv} the occurrence of spontaneous stable vortices in
OPO is described for clean cavities, while the case of disordered
cavities is studied at the end of Sec.~\ref{sec:numer}. Next we
describe in general terms the role of adding a pulsed Gaussian probe
to the OPO regime (the so called TOPO regime) in Sec.~\ref{sec:topor},
while metastable vortices triggered by a Laguerre-Gauss probe are
discussed in Sec.~\ref{sec:trigv}. Here, in Sec.~\ref{sec:onset}, we
also describe the onset and dynamics of vortex-antivortex
pairs. Stability of multiply quantised vortices is analysed in
Sec.~\ref{sec:stabi} and finally we mention the occurrence of vortices
in polariton fluids in other regimes than OPO in Sec.~\ref{sec:other}.


\section{A very short introduction to microcavity polaritons}
\label{sec:intpo}
Before focusing on the main topic of this review, we give here a very
short introduction to microcavity polaritons in order to fix the
notation for later on. A more complete introduction can be found in
several review
articles~\cite{skolnick98,savona99,khitrova99,ciuti03,keeling_review07,deng10}
and books~\cite{yamamoto,yamamoto00,kavokin-book,kavokin_laussy} on
microcavity polaritons.

Microcavity polaritons are the normal modes resulting from the strong
coupling between quantum well (QW) excitons and cavity photons. In
semiconductor microcavities, the mirrors employed to confine light are
distributed Bragg reflectors, i.e., alternating quarter wavelength
thick layers of dielectrics with different refractive indices. Between
the Bragg reflectors, the cavity light forms a standing wave pattern
of confined radiation, which can be described by an approximatively
quadratic dispersion, $\omega_C(k) = \omega_C^0 + k^2/(2m_C)$ (from
here onwards we fix $\hbar = 1$). 
Excitons are the hydrogenic bound states of a conductance band
electron and a valence band hole, therefore their mass is much larger
than the cavity photon mass (typically $m_C\sim 10^{-5} m_e$, where
$m_e$ is the free electron mass). For this reason, exciton dispersion
can be neglected, $\omega_X(k) = \omega_X^0$. In microcavities, one or
multiple QWs are grown in between the mirrors, so that excitons are at
the antinodes of the confined light, giving rise to strong coupling.
In addition, cavity mirrors are built with a wedge, so as to change
the detuning between the normal incidence energy of the cavity field
and the exciton one, $\delta \equiv \omega_C^0 - \omega_X^0$. Typical
parameter values for a GaAs-based microcavity are listed in
Tab.~\ref{tab:value}.

\begin{figure}
\begin{center}
\includegraphics[width=0.7\linewidth,angle=0]{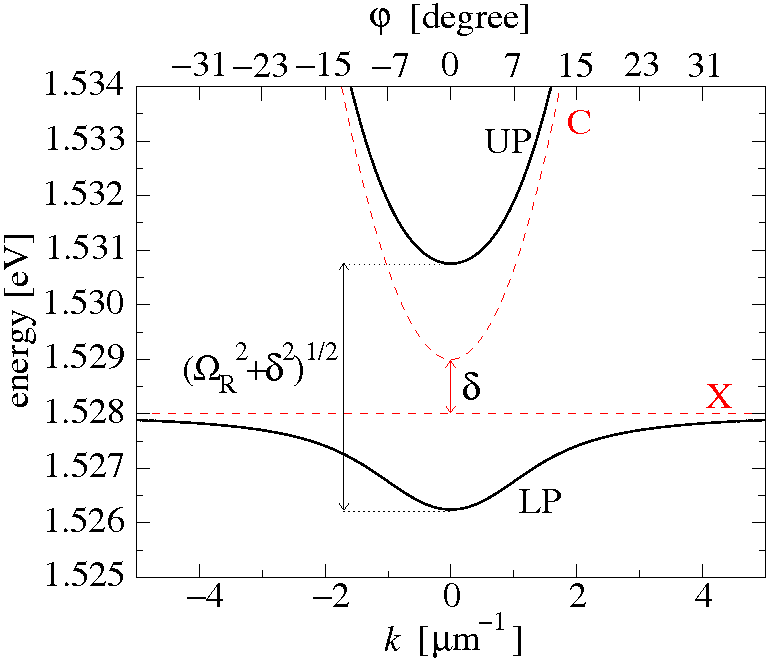}
\end{center}
\caption{Lower (LP) and upper polariton (UP) energy dispersions (solid
  black), together with the dispersions of the photon (C) and exciton
  (X) fields (red dashed) as a function of either the wave-vector
  $k~[\mu\text{m}^{-1}]$ or the emission angle $\varphi$~[degree] for
  $m_C=2.3\times 10^{-5}$, $\Omega_R=4.4$~meV, $\omega_X^0 =
  1.5280$~eV, and a detuning $\delta = 1$~meV.}
\label{fig:bared}
\end{figure}
The polariton normal modes can be found by solving the coupled
Schr\"odinger equations for exciton and photon fields, $\psi_{X,C} =
\psi_{X,C} (\vect{r},t)$
\begin{align}
  i\partial_t \begin{pmatrix} \psi_X \\ \psi_C \end{pmatrix} &=
  \hat{H}_0 \begin{pmatrix} \psi_X \\ \psi_C \end{pmatrix} & \hat{H}_0
  &= \begin{pmatrix} \omega_{X}^0 - i \kappa_X & \Omega_R/2
    \\ \Omega_R/2 & \omega_C (-i\nabla) - i \kappa_C \end{pmatrix} \; ,
\label{eq:polar}
\end{align}
where $\Omega_R$ is the Rabi splitting and $\kappa_{X,C}$ are the
decay rates of exciton and photon. For an ideal cavity,
$\kappa_{X,C}=0$, the eigenstates of this equations in momentum space,
$\psi_{X,C}(\vect{r},t) = e^{i\omega t} \sum_{\vect{k}}
e^{i\vect{k}\cdot \vect{r}} \psi_{X,C,\vect{k}}$, are the lower (LP)
and upper polaritons (UP)
\begin{align}
\label{eq:rotlp}
  \begin{pmatrix} \psi_{X,\vect{k}} \\ \psi_{C,\vect{k}} \end{pmatrix}
  &= \begin{pmatrix} \cos\theta_{k} & -\sin\theta_{k}
    \\ \sin\theta_{k} & \cos\theta_{k} \end{pmatrix}\begin{pmatrix}
    \psi_{LP,\vect{k}} \\ \psi_{UP,\vect{k}} \end{pmatrix}\\
  \cos^2\theta_{k}, \sin^2 \theta_{k} &= \Frac{1}{2} \left(1 \pm
  \Frac{\omega_C(k) - \omega_X^0}{\sqrt{(\omega_C(k) - \omega_X^0)^2 +
      \Omega_R^2}}\right)\; ,
\end{align}
with an energy dispersion given by (see Fig.~\ref{fig:bared}):
\begin{equation}
  \omega_{LP,UP} (k) = \Frac{1}{2} \left[\omega_C(k) +
    \omega_X^0\right] \mp \Frac{1}{2} \sqrt{\left[\omega_C(k) -
      \omega_X^0\right]^2 + \Omega_R^2} \; .
\label{eq:bared}
\end{equation}
At zero detuning ($\delta=0$) and normal incidence ($k=0$) polaritons
are exactly half-light and half-matter quasi-particles ($\cos^2
\theta_0 = 0.5 = \sin^2 \theta_0$).
The value of the momentum $k$ of polaritons inside the cavity is
related to the emission angle $\varphi$ (with respect to normal
incidence) of photons outside the cavity by $ck=\omega_{LP} (\vect{k})
\sin\varphi$. Thanks to this property, microcavity polaritons can be
directly excited by a laser field and detected via reflection,
transmission or photoluminescence measurements. In
Fig.~\ref{fig:bared}, the energy dispersion of the lower and upper
polariton are plotted as a function of both wave-vector,
$k~[\mu\text{m}^{-1}]$, or the emission angle, $\varphi$~[degree], for
typical values of microcavity parameters.

\subsection{Exciton-exciton and exciton-photon interaction}
A fundamental property of polaritons is their non-linear behaviour
inherited from the exciton-exciton interaction and the saturation of
the exciton-photon coupling. In this review, we treat excitons as
bosonic particles, therefore the effective exciton-exciton interaction
can be written as
\begin{equation*}
  \mathcal{H}_{XX} = \frac{1}{2 A} \sum_{\vect{k},\vect{k}',\vect{q}}
  V_{q} \psi_{X,\vect{k}+\vect{q}}^* \psi_{X,\vect{k}'-\vect{q}}^*
  \psi_{X,\vect{k}}^{\phantom{*}} \psi_{X,\vect{k}'}^{\phantom{*}} \; ,
\end{equation*}
where the effective interaction potential $V_{q}$ can be determined
starting from the microscopic electron-hole
Hamiltonian~\cite{ciuti98:prb,rochat00:prb}. The typical wave-vectors
involved in the physics described by this review are much smaller than
the inverse exciton Bohr radius, $q \ll a_X^{-1}$, where $a_X =
\epsilon/(2 \mu e^2)$ is the two-dimensional exciton Bohr radius,
$\epsilon$ the dielectric constant, and $\mu^{-1} = m_e^{-1} +
m_h^{-1}$ the electron-hole reduced mass. In this limit, it can be
shown~\cite{ciuti98:prb} that the momentum dependence of $V_{q}$ can
be neglected, thus approximating it with a contact interaction, $V_{q}
\simeq g_X = 6e^2 a_X/\epsilon = 6 \Ry_X a_X^2$, where $\Ry_X =
e^2/(\epsilon a_X) = 1/(2\mu a_X^2)$ is the exciton
Rydberg. Typically, for GaAs quantum wells (see Tab.~\ref{tab:value}),
$\epsilon = 13$, $a_X \simeq 7$~nm, and $\Ry_X \simeq 17$~meV ,
therefore $g_X \simeq 0.005$~meV$(\mu\text{m})^2$. We will see,
however, that the exact value of the coupling constant $g_X$ has no
relevance for the mean-field dynamics we are going to describe, i.e.,
$g_X$ can be rescaled to $1$.
\begin{table}
\begin{center}
\begin{tabular}{|c l|c l|}
\hline\hline
\quad QW \qquad & & \quad cavity \qquad & \\
\hline
\qquad & $\epsilon \simeq 13$ & \quad & $\omega_C^0 \simeq \omega_X^0 \simeq 1.53$~eV\\
\quad & $m_e = 0.063m^0_e$    & \quad & $\delta \in [-10,10]$~meV\\
\quad & $m_h = 0.3m^0_e$      & \quad & $m_C = 2.3\times 10^{-5} m^0_e$ \\
\quad & $a_X \simeq 70~\AA$   & \quad & $\ell_C = 0.868~ \mu$m\\
\quad & $\Ry_X \simeq 17$~meV  & \quad & $\Omega_R \simeq 4.4$~meV \\
\quad & $\kappa_X \sim \mu$eV     & \quad & $\kappa_C = 0.1$~meV \\
\hline\hline
\end{tabular}
\end{center}
\caption{Characteristic parameters of a GaAs-based microcavity,
  divided between the parameters of the quantum well (left) and those
  describing the microcavity (right). Here, $\ell_C=\sqrt{1/(m_C 
    \Omega_R)}$ is a characteristic length for the cavity photons. The
  photon decay rate $\kappa_C$ refers to a cavity mirror with
  typically 25 bottom pairs and 15 lower pairs (see, e.g.,
  Ref.~\cite{sanvitto10}).} 
\label{tab:value}
\end{table}

The composite nature of excitons, as a bound state of an electron and
a hole, is also visible in the saturability of the exciton-photon
coupling, resulting in an anharmonic interaction term which adds
to the usual harmonic one:
\begin{equation}
  \mathcal{H}_{XC} = \frac{\Omega_R}{2}\int d\vect{r}
  \left[\psi_X^*(\vect{r})\psi_C (\vect{r}) + \psi_C^*(\vect{r})\psi_X
    (\vect{r})\right] \left[1 -
    \frac{|\psi_X(\vect{r})|^2}{n_{\text{sat}}}\right]\; ,
\label{eq:satur}
\end{equation}
where $n_{\text{sat}} = 7/(16 \pi a_X^2)$ is the exciton saturation
density~\cite{rochat00:prb}. In GaAs, $n_{\text{sat}}\simeq
2842$~$(\mu\text{m})^{-2}$ and for a Rabi splitting of
$\Omega_R=4.4$~meV, the ratio between saturation and exciton-exciton
interaction strength,
\begin{equation*}
  \frac{\Omega_R}{2 g_X n_{\text{sat}}} \simeq 0.1\; ,
\end{equation*}
allow us to neglect the anharmonic term in~\eqref{eq:satur} for the
kind of physics we want to describe in this review.

Therefore, the mean-field evolution of the coupled cavity
photon--exciton dynamics is described by the following non-linear
Schr\"odinger equation or Gross-Pitaevskii equation (GPE):
\begin{equation}
  i\partial_t \begin{pmatrix} \psi_X \\ \psi_C \end{pmatrix} =
  \left[\hat{H}_0 + \begin{pmatrix} g_X |\psi_X|^2 & 0\\ 0 &
      V_C(\vect{r})\end{pmatrix}\right] \begin{pmatrix} \psi_X
    \\ \psi_C \end{pmatrix} \; .
\label{eq:polai}
\end{equation}
Here, we have also added an external potential $V_C(\vect{r})$ acting
on the photon component, which later on we will use to describe the
effect of photonic disorder present in the cavity mirrors.
Note that Eq.~\eqref{eq:polai} is a classical field description, which
assumes the macroscopic occupation of a finite number of states, each
described by a complex classical function $\psi$.

\section{Optical parametric oscillator regime}
\label{sec:opore}
An accurate control of the polariton dynamics can be achieved by
directly injecting polaritons at a given wave-vector and frequency
with a properly tuned external laser --- the resonant excitation
scheme.
In within this scheme two regimes can be singled out: (i) the regime
where only the polariton state generated by the pump is a stable
configuration of the system (we refer to this as the \emph{pump-only}
state); (ii) the regime where the polaritons continuously injected
into the \emph{pump} state undergo
coherent stimulated scattering into the \emph{signal} state (close to
the normal direction) and the \emph{idler} state (on the other side of
the pump). 
Parametric scattering from pump to signal and idler can be
self-induced by the continuous-wave (cw) laser above a pump strength
threshold, in which case one refers to the optical parametric
oscillator (OPO) regime. However, below the threshold for OPO, a
second weak probe beam shined close to either the (expected) signal or
idler states, can be used to `seed' the parametric scattering
processes and amplify the probe; in this case, one refers to the
optical parametric amplification (OPA) regime.

We introduce the concept of polariton parametric scattering and review
the main experimental results on optical parametric amplification in
the next section. In Sec.~\ref{sec:plane}, we use a simplified
theoretical model in terms of plane waves, for both the pump-only
resonant state and the OPO state, summarising the main properties of
both regimes and drawing an analogy with equilibrium weakly
interacting Bose-Einstein condensates (BECs). Finally, in
Sec.~\ref{sec:numer}, we explain the necessity for carrying out a
numerical analysis of the OPO.
%
Much experimental work has been carried out on polaritons in the OPO
regime~\cite{stevenson00,houdre00,baumberg00:prb,tartakovskii02,krizhanovskii02,butte03,gippius04,baas06:prl,sanvitto06,krizhanovskii06prl,krizhanovskii08,ballarini09}
(for a review on the experiments, see Ref.~\cite{skolnick03}). We will
discuss the experimental achievements along with the theoretical
description.

\subsection{Polariton parametric scattering and optical parametric
  amplification}
\label{sec:ampli}
\begin{figure}
\begin{center}
\includegraphics[width=1.0\linewidth,angle=0]{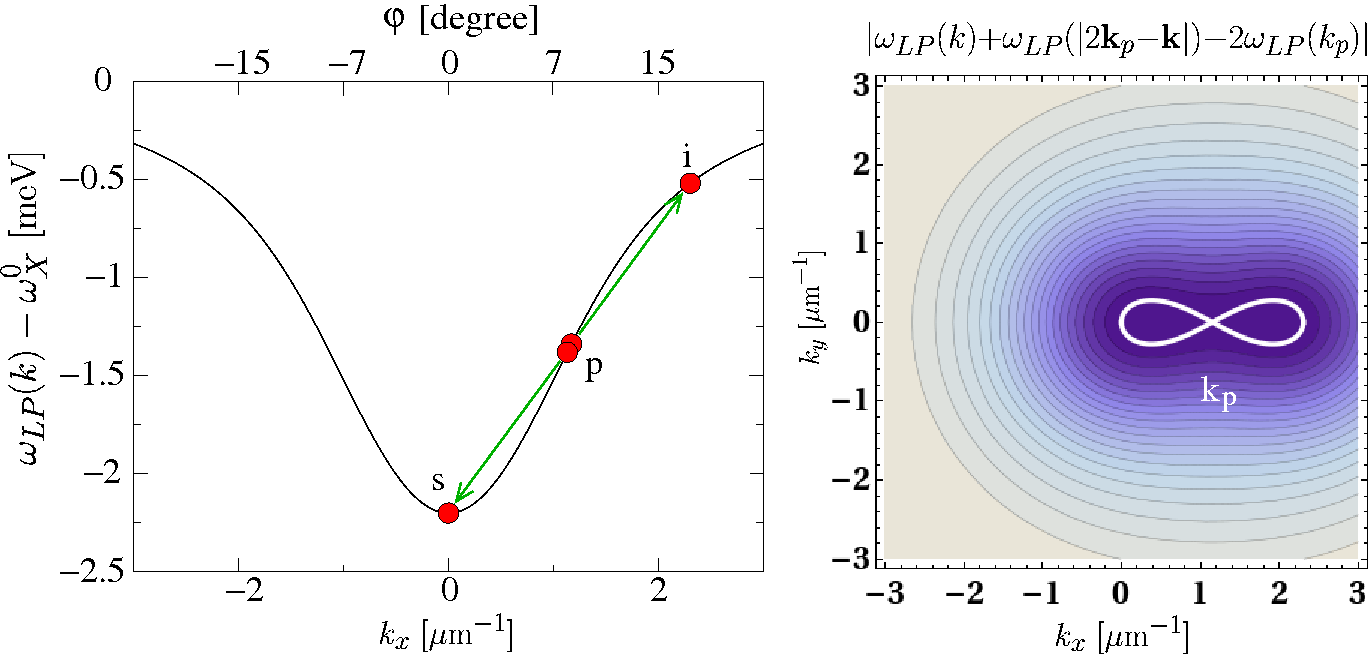}
\end{center}
\caption{Illustration of the basic idea of parametric
  scattering. Left: two LPs scatter from the pump state
  $\{\vect{k}_p,\omega_p\}$ towards the signal
  $\{\vect{k}_s,\omega_s\}$ (here at zero momentum) and the idler
  state $\{\vect{k}_i = 2\vect{k}_p - \vect{k}_s, \omega_i = 2\omega_p
  - \omega_s\}$ (at higher momentum), conserving momentum and
  energy. Right: Following Refs.~\cite{ciuti01,ciuti03}, we plot
  $|\omega_{LP} (k)+\omega_{LP}
  (|2\vect{k}_p-\vect{k}|)-2\omega_{LP}(k_p)|$ as a function of
  $\vect{k}=(k_x,k_y)$. The white line is the zero value of the
  contour and the pump is oriented along the $x$-direction,
  $\vect{k}_p=(k_p,0)$. The parameters used in both panels are the
  same as the ones of Fig.~\ref{fig:bared}.}
\label{fig:param}
\end{figure}

In the parametric scattering process, two polaritons from a pump mode,
with wave-vector and frequency $\{\vect{k}_p,\omega_p\}$, scatter into a
lower energy signal mode $\{\vect{k}_s,\omega_s\}$ and a higher energy
idler mode $\{\vect{k}_i,\omega_i\}$. This scattering process has to conserve energy
and momentum, therefore requiring that
\begin{align}
  2 \vect{k}_p &= \vect{k}_s + \vect{k}_i & 2\omega_p &= \omega_s +
  \omega_i\; .
\label{eq:match}
\end{align}
%
This condition cannot be satisfied by any particle dispersion, for
example parametric scattering is forbidden for particles with a
quadratic dispersion. In order to check whether parametric scattering
is allowed for polaritons, one has to verify if the condition
\begin{equation}
  2\omega_{LP} (k_p) = \omega_{LP} (k_s) + \omega_{LP} (|2\vect{k}_p -
  \vect{k}_s|)
\end{equation}
can be satisfied. If $\vect{k}_s=0$, then the momenta of pump and
idler are uniquely selected (see left panel of
Fig.~\ref{fig:param}). In this case, the value of the pumping angle is
also referred to as the ``magic angle'', and is located close to the
inflection point of the LP dispersion. However, for a generic signal
wave-vector, $\vect{k}_s=\vect{k}=(k_x,k_y)$, then for a fixed pump
angle $\vect{k}_p$ (assumed to be oriented along the $x$-direction,
$(k_p,0)$, in the right panel of Fig.~\ref{fig:param}), the final
states allowed in the parametric scattering process describe a
figure-of-eight in momentum space~\cite{ciuti01,ciuti03,langbein04}.

In the case of optical parametric amplification experiments,
parametric scattering is stimulated by a weak additional probe
field. OPA was first observed in an InGaAs/GaAs/AlGaAs
microcavity~\cite{savvidis00:prl}, where a substantial signal gain of
up to 70 was measured. Much experimental work has followed this first
result~\cite{savvidis00:prb,huang00,dasbach2000,erland2000,messin01,saba01,savvidis01,kundermann03,huynh03,diederichs06}.
Pump-probe parametric amplification of polaritons with an
extraordinary gain up to 5000 and at temperatures up to 120K has been
reached in GaAlAs-based microcavities and up to 220K in CdTe-based
microcavities~\cite{saba01}. In three-beam pulsed
experiments~\cite{huang00}, polaritons scatter from two equal and
opposite angles, $\vect{k}_p$ and $-\vect{k}_p$, into the LP and UP
states at $\vect{k}=0$ --- note that at zero detuning, $\delta=0$,
$2\omega_X^0 = \omega_{LP} (0) + \omega_{UP} (0)$.
%
%
Interestingly, parametric amplification has been also obtained for
ultracold atom pairs confined in a moving one-dimensional optical
lattice~\cite{campbell06}. The role of the periodic optical lattice is
to deform the atom dispersion from the quadratic one, allowing
parametric scattering to happen.

The stimulated scattering regime can be reached also in the OPO
configuration, i.e., without an additional probe beam. Now, stimulated
scattering is self-initiated at pump powers above a threshold
intensity, where the final state population is close to one. We will
see that in this case, there is no special significance of the ``magic
angle'', rather a broad range of pumping angles larger than a critical
value, ($\theta_p \gtrsim 10^{\circ}$ for the parameters in
Ref.~\cite{whittaker05}) allow OPO with a signal emission close to
normal incidence, $\theta_s \simeq 0^{\circ}$. In addition, for finite
size pumping (see later Sec.~\ref{sec:numer}), the pump, signal, and
idler momenta are smeared in a broad interval, while their frequency
still satisfy the matching conditions~\eqref{eq:match} exactly.
In the next three sections, we will focus mainly on the theoretical
description of polariton resonant excitation with a cw laser field,
describing the properties of first the pump-only state and then the
OPO state.

\subsection{Bistability and OPO in the plane-wave approximation}
\label{sec:plane}
The theoretical description of polaritons in the resonant excitation
regime can be formulated in terms of the same classical two-field
non-linear Schr\"odinger equation previously introduced in
Eq.~\eqref{eq:polai}, where now an external driving field
$F_p(\vect{r},t)$ is added in order to describe the coherent injection
of photons into the cavity:
\begin{equation}
  i\partial_t \begin{pmatrix} \psi_X \\ \psi_C \end{pmatrix}
  = \begin{pmatrix} 0\\ F_p(\vect{r},t) \end{pmatrix} + \left[\hat{H}_0
    + \begin{pmatrix} g_X |\psi_X|^2 & 0\\ 0 &
      V_C(\vect{r})\end{pmatrix}\right] \begin{pmatrix} \psi_X
    \\ \psi_C \end{pmatrix}\; .
\label{eq:ppump}
\end{equation}
A continuous-wave (cw) pumping laser can be written as
\begin{equation}
  F_p(\vect{r},t) = \mathcal{F}_{f_p,\sigma_p} (r) e^{i (\vect{k}_p
    \cdot \vect{r} - \omega_p t)} \; ,
\label{eq:pumpe}
\end{equation}
where $\mathcal{F}_{f_p,\sigma_p} (r)$ can either describe a
homogeneous pump with strength $f_p$, $\mathcal{F}_{f_p,\sigma_p}(r) =
f_p$, or, as we will assume later in Sec.~\ref{sec:numer}, a Gaussian
or a top-hat spatial profile with strength $f_p$ and full width at
half maximum (FWHM) $\sigma_p$.

For a homogeneous pump, $\mathcal{F}_{f_p,\sigma_p} (r) = f_p$, and
for a clean system, $V_C(\vect{r}) = 0$, the conditions under which a
stable OPO switches on can be found by making use of an analytical
treatment~\cite{ciuti03,whittaker05,ciuti05,wouters07:prb}. In fact,
in this limit, each mode can be approximated as a plane wave.
To simplify the analytical expressions, it is useful to rotate
Eq.~\eqref{eq:ppump} into the LP and UP basis, as described by
Eq.~\eqref{eq:rotlp}, and to neglect the contribution from the UP
states, assuming that the LP and UP branches are not mixed together by
the non-linear terms.
In this case, working in momentum space, $\psi_{LP}(\vect{r},t) =
\sum_{\vect{k}} e^{i\vect{k}\cdot \vect{r}} \psi_{LP,\vect{k}} (t)$,
Eq.~\eqref{eq:ppump} can be written as
\begin{multline}
  i\partial_t \psi_{LP,\vect{k}} = \left[\omega_{LP} (k) -
    i\kappa (k)\right]\psi_{LP,\vect{k}} +\\
  \sum_{\vect{k}_1, \vect{k}_2} g_{\vect{k}, \vect{k}_1, \vect{k}_2}
  \psi^*_{LP,\vect{k}_1 + \vect{k}_2-\vect{k}} \psi_{LP,\vect{k}_1}
  \psi_{LP,\vect{k}_2} + \sin\theta_k f_p e^{-i\omega_p t}
  \delta_{\vect{k},\vect{k}_p}\; ,
\label{eq:efflp}
\end{multline}
where $\kappa(k)=\kappa_X \cos^2\theta_k + \kappa_C \sin^2\theta_k$ is
the effective LP decay rate and the interaction strength now reads as
$g_{\vect{k}, \vect{k}_1, \vect{k}_2}=g_X \cos\theta_{k}
\cos\theta_{|\vect{k}_1 + \vect{k}_2-\vect{k}|} \cos\theta_{k_1}
\cos\theta_{k_2}$.

If we consider solutions of the Eq.~\eqref{eq:efflp} where only the
pump mode, $\vect{k}=\vect{k}_p$, is populated, we can find an exact
solution in the form
\begin{align}
  \psi_{LP}(\vect{r},t) & =p e^{i(\vect{k}_p \cdot \vect{r} -
    \omega_pt)} & \psi_{LP,\vect{k}} (t) &= p
  \delta_{\vect{k},\vect{k}_p} e^{-i\omega_p t}\; ,
\label{eq:pumpo}
\end{align}
where the complex amplitude $p$ is given by
\begin{equation}
  \left[\omega_{LP}(k_p) -\omega_p - i \kappa(k_p) + g_X
    \cos^4\theta_{k_p} |p|^2\right] p + \sin\theta_{k_p} f_p = 0\; .
\label{eq:cubic}
\end{equation}
For practical purposes, one can substitute $g_X \cos^4\theta_{k_p}
\mapsto 1$ by redefining the pump strength $f_p' = \sqrt{g_X}
\cos^2\theta_{k_p} \sin\theta_{k_p} f_p$, and rescaling the field
strength $p$ by $|p'|=\sqrt{g_X} \cos^2\theta_{k_p} |p|$. Note that
the $\chi^{(3)}$-non-linear interaction term, $|p'|^2p'$, renormalises
the effective detuning of the pump mode from the LP dispersion,
\begin{equation}
 \Delta_p \equiv \omega_p-\omega_{LP} (k_p)-|p'|^2\; ,
\label{eq:detun}
\end{equation}
which now includes the blue-shift of the LP dispersion due to
interactions.

\begin{figure}
\begin{center}
\includegraphics[width=1.0\linewidth,angle=0]{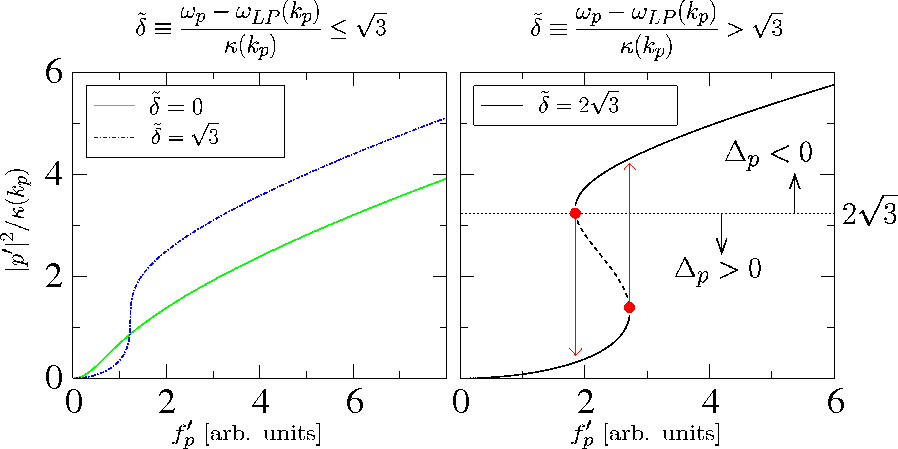}
\end{center}
\caption{Dimensionless LP population $|p'|^2/\kappa(k_p) = g_X \cos^4
  \theta_{k_p} |p|^2 /\kappa(k_p)$ as a function of the dimensionless
  pump intensity $f_p'=\sqrt{g_X} \cos^2\theta_{k_p} \sin\theta_{k_p}
  f_p$ for different values of the parameter $\tilde{\delta} =
  [\omega_p - \omega_{LP} (k_p)]/\kappa(k_p)$. When $\tilde{\delta}
  \le \sqrt{3}$ (left panel), the system is in the optical limiter
  regime, while for sufficiently blue-detuned pump frequencies,
  $\tilde{\delta} > \sqrt{3}$, a bistable behaviour is obtained
  (right). The sign of the interaction renormalised detuning,
  $\Delta_p$~\eqref{eq:detun}, is also given.}
\label{fig:bista}
\end{figure}
The general solution of the cubic equation~\eqref{eq:cubic} is well
known, 
and exhibits a qualitatively different behaviour depending whether the
pump frequency is blue- or red-detuned with respect to the LP
dispersion. In particular, if $\omega_p-\omega_{LP} (k_p) \le \sqrt{3}
\kappa(k_p)$, the system is in the optical limiter regime, where the
population $|p'|^2$ grows monotonically as a function of the pump
intensity $f_p'$. If instead $\omega_p-\omega_{LP} (k_p) > \sqrt{3}
\kappa(k_p)$, the system displays bistable behaviour, with a
characteristic $S$-shape of $|p'|^2$ as a function of $f_p'$, the
second turning point coinciding with the point where the effective
detuning $\Delta_p$~\eqref{eq:detun} changes sign (see
Fig.~\ref{fig:bista}).
Because the branch with negative slope is unstable, the polariton
density in the pump-only mode follows a hysteretic behaviour:
Increasing the pump intensity, eventually the pump-only mode jumps
abruptly into the upper branch, while if the intensity is then
decreased, the polariton population decreases and jumps back down to
the lower branch for smaller values of the pump intensity. Optical
bistability in microcavity polaritons has been observed in
Refs.~\cite{baas04,cavigli05}, with evidence of a hysteresis cycle of
the polariton emission as a function of the pump intensity.
Multistability of two different polariton states, generated by either
populating two different spin
states~\cite{gippius07,adrados10,paraiso10} or by injecting two states
with two different pumps~\cite{cancellieri11} has been also recently
proposed and, in the spin case, observed.
Part of the interest in this field is to realise all-optical switches
~\cite{amo10} and memories.

The dynamical stability of the pump-only solution~\eqref{eq:pumpo} can
be established by allowing other states than the pump (i.e., the
signal and idler states) to be perturbatively populated via parametric
scattering processes, 
\begin{equation}
  \psi_{LP,\vect{k}} (t) = p \delta_{\vect{k},\vect{k}_p}
  e^{-i\omega_p t} + s \delta_{\vect{k},\vect{k}_p-\vect{q}}
  e^{-i(\omega_p-\omega) t} + i^*
  \delta_{\vect{k},\vect{k}_p+\vect{q}} e^{-i(\omega_p+\omega^*) t}\;
  ,
\label{eq:liopo}
\end{equation}
where, $\{\vect{k}_{s,i} = \vect{k}_p\mp \vect{q},\omega_{s,i} =
\omega_p \mp \omega\}$, and by assessing whether the time evolution of
these states grows exponentially in time or not. Expanding to the
first order in $s$ and $i$, one obtains an eigenproblem for the
amplitudes $s$ and $i$~\cite{whittaker05,ciuti05,wouters07:prb},
\begin{equation}
  \begin{pmatrix} \omega -\Delta_s - i \kappa(k_s) & g_X c_s c_i c^2_p
    p^2\\ g_X c_s c_i c^2_p {p^*}^2 & -\omega -\Delta_i + i
    \kappa(k_i) \end{pmatrix} \begin{pmatrix} s \\ i\end{pmatrix} = 0
    \: ,
\label{eq:mbogo}
\end{equation}
where $\Delta_{s,i} = \omega_{p} - \omega_{LP}(k_{s,i}) - 2 g_X
c_{s,i}^2 c_p^2 |p|^2$ and $c_{p,s,i}=\cos\theta_{k_p,k_s,k_i}$. The
complex eigenvalues $\omega$ can be obtained imposing that the
determinant of the matrix in~\eqref{eq:mbogo} is zero. The dynamical
stability is ensured if $\Im(\omega)>0$. Therefore, the threshold for
instability of the pump-only solution~\eqref{eq:pumpo} can be found
imposing the condition $\Im(\omega)=0$. By fixing the pump wave-vector
and energy $(\vect{k}_p,\omega_P)$ and the signal wave-vector
$\vect{k}_s$ (as well as the exciton and photon lifetimes,
$\kappa_{X,C}$), this provides a criterion for establishing the
boundaries of the instability region, i.e., the lowest and highest
values of the LP population $|p|^2$ for which the pump-only solution
is not stable.
As shown in Refs.~\cite{whittaker05,ciuti05,wouters07:prb}, one can
classify the instability as a single mode instability when
$\vect{q}=0$ and therefore $\vect{k}_p=\vect{k}_s=\vect{k}_i$ --- the
Kerr instability. In particular, the branch with negative slope of the
bistable curve (dashed line in Fig.~\ref{fig:bista}) is always single
mode unstable~\cite{whittaker05,wouters07:prb}. If instead
$\vect{q}\ne 0$, the instability is parametric-like. Now, the total
extent of the instability region corresponding to different values of
$\vect{k}_s$ is significantly larger than just the branch with
negative slope. In addition, the OPO state does not require a bistable
behaviour and can turn on also in the optical limiter case.
In particular, it is possible to plot a ``phase
diagram''~\cite{whittaker05} of pump energy $\omega_P$ as function of
pump wave-vector $\vect{k}_p$, showing the regions where a pump-only
solution is always stable, where the OPO switches on, and where
instead a Kerr-type instability is only possible.
In this way, in Ref.~\cite{whittaker05}, it was shown that there is no
particular significance to the ``magic angle'' for the pump. Rather,
OPO conditions can be found for all angles larger than a critical
value, $\theta_p>\theta_c$ ($\sim 10^{\circ}$ for the parameters of
Ref.~\cite{whittaker05}), as also confirmed
experimentally~\cite{butte03,gippius04}. In addition, the energy
renormalisation of the polariton dispersion due to interactions moves
the emission angles for the signal always close to $\theta_s \sim
0$~\cite{whittaker05,gippius04}. This is also confirmed by the
numerical simulations we have carried out and illustrated later on in
Sec.~\ref{sec:numer}.

The method described above implies negligible populations of the
signal $s$ and the idler $i$ and therefore it allows to find the
conditions for the OPO \emph{threshold}. In order to find the OPO
states, one cannot linearise in $s$ and $i$, but instead include the
contributions of finite signal and idler populations to the dispersion
renormalization~\cite{whittaker05}.
%
In this way, in the region unstable for parametric scattering
determined with the method described above, one can describe first the
increase (switch-on) and later the decrease (switch-off) of the signal
and idler populations as a function of the pump power.
It is interesting to note that by doing that, i.e., by
substituting~\eqref{eq:liopo} into~\eqref{eq:efflp}, ``satellite
states'' oscillating with energies
$\omega_{s_2}=2\omega_s-\omega_p=\omega_p-2\omega$ and
$\omega_{i_2}=2\omega_i-\omega_p=\omega_p+2\omega$ automatically
appear. In fact, above OPO threshold, when signal and idler
populations are not negligible, parametric scattering from the signal
(idler) state into the pump and second-signal (second-idler) satellite
state take place, i.e., $2s \mapsto p + s_2$ ($2i \mapsto p + i_2$)
and therefore $2\omega_s = \omega_p + \omega_{s_2}$ ($2\omega_i =
\omega_p + \omega_{i_2}$).
This is clearly seen in the
``exact'' OPO solution obtained numerically (see, .e.g.,
Fig.~\ref{fig:spect}), as well as it has been observed experimentally
(see, e.g., Ref.~\cite{tartakovskii02}). One has to note however that
the population of the ``satellite states'' by multiple scattering
processes is always negligible w.r.t. that one of pump, signal, and
idler (see right panel of Fig.~\ref{fig:spect}).

We will introduce the numerical modelling used to describe the problem
for a finite-size pump later in Sec.~\ref{sec:numer}. Before doing
that, in the next section , we concentrate on the analogies and
differences between an OPO state and an equilibrium weakly interacting
Bose-Einstein condensate (BEC).

\subsubsection{Spontaneous $U(1)$ phase symmetry breaking and
  Goldstone mode}
\label{sec:golds}
The OPO state looks at first sight very different from an equilibrium
weakly interacting BEC. In particular, the OPO is an intrinsically
non-equilibrium state characterised by the (macroscopic) occupation of
three polariton states only, one directly populated by the external
pump and the signal and idler states populated by parametric
scattering. Contrast this with the thermodynamic phase transition in a
BEC, where the macroscopic occupation of the ground state occurs when,
for a thermal distribution of bosons, either the temperature is
lowered below a critical value or the density is increased.
The OPO state does however share with a BEC the fundamental property
of spontaneous symmetry breaking of the phase
symmetry~\cite{whittaker05,wouters06b}. In fact, the external laser
fixes the phase of the pump state $\phi_p$ and parametric scattering
processes constraint the sum of the signal and the idler phase only,
$2\phi_p=\phi_s + \phi_i$, but leaves the system to arbitrarily choose
the phase difference $\phi_s - \phi_i$. In other words, one can easily
show that the system of three equations one obtains by imposing the
OPO solution~\eqref{eq:liopo} into the mean-field
equation~\eqref{eq:efflp} is invariant for a simultaneous phase
rotation of both signal and idler states:
\begin{align}
  s &\mapsto se^{i \phi} & i &\mapsto ie^{-i \phi} \; .
\end{align}
This $U(1)$ phase rotation symmetry gets spontaneously broken in the
OPO regime, where the signal and idler spontaneously select their
phase, though not independently. Note that in this respect the OPO
regime differs very much from the optical parametric amplification
(OPA) regime, where both signal and idler phases are fixed by the
external probe, and therefore the $U(1)$ phase rotation symmetry is
explicitly broken by the probe and no phase freedom is left in the
system.

Goldstone's theorem states that the spontaneous symmetry breaking of
the $U(1)$ phase symmetry in OPO is accompanied by the appearance of a
gapless soft mode, i.e., a mode $\omega(\vect{k})$ whose both
frequency $\Re[\omega (\vect{k})]$ and decay rate
$\Im[\omega(\vect{k})]$ tend to zero in the long wave-length $\vect{k}
\to 0$ limit. The dispersion for the Goldstone mode in OPO has been
derived in Ref.~\cite{wouters06b}, where also an experimental set-up
to probe it's dispersion has been proposed. 
In addition, the appearance of spontaneous coherence in OPO have been
shown via quantum Monte Carlo simulations~\cite{carusotto05} through
the divergence of the coherence length when the pump intensity
approaches the threshold.
In contrast, in the OPA regime, where the phase rotation symmetry is
explicitly broken by the probe, there is no Goldstone mode and a gap
opens in the imaginary part of the elementary excitation dispersion,
$\Im[\omega(\vect{k})]$.


We would like to stress here that, even though an equilibrium weakly
interacting BEC and an OPO state share the fundamental property of
spontaneous symmetry breaking of the phase symmetry, some care needs
to be applied in pushing this analogy further. In particular, the
existence of a free phase alone is not sufficient to ensure the
paradigmatic properties of a superfluid, such as the Landau criterion,
the stability of quantised vortices, and the persistency of metastable
flow. For example, let us consider here the case of the Landau
criterion: In an equilibrium weakly interacting BEC, the existence of
the soft Goldstone mode (the Bogoliubov mode), with it's
characteristic linear dispersion for $\vect{k} \to 0$,
$\omega(\vect{k}) \simeq c_s k$ , implies the existence of a critical
velocity, $v_c \equiv \text{min}_\vect{k} \omega(\vect{k})/k = c_s$
(the speed of sound), below which a perturbative defect dragged
through the fluid cannot dissipate energy (superfluid regime).  In the
non-equilibrium OPO regime instead, similarly to what happens for
incoherently pumped polaritons
condensates~\cite{szymanska06:prl,wouters:140402,wouters10}, the
unusual form of the excitation spectrum --- diffusive at small momenta
--- poses fundamental questions on the fulfilling of the Landau
criterion and the possibility of dissipationless superflow.

Similarly, properties such as the appearance and stability of
quantised vortices and the persistency of metastable flow need to be
independently assessed in polariton fluids in the three different
pumping schemes available --- (i) non-resonant pumping; (ii)
parametric drive in the optical-parametric-oscillator regime; (iii)
coherent drive in the pump-only configuration. In fact, in the case of
an equilibrium condensate, the ground state is flow-less, i.e. a
vortex solution is unstable in non rotating condensates~\footnote{\ In
  rotating condensates, a vortex can be created if the angular
  velocity is higher than a critical
  value~\cite{hess67,pitaevskii03}. When rotation is halted, then the
  vortex will spiral out of the condensate~\cite{rokhsar97}.}. In
contrast, in a polariton fluid, its intrinsic non-equilibrium nature
implies the presence of a flow even when a steady state regime is
reached. In this sense, not always the presence of vortices can be
ascribed to the superfluid property of the system. We will discuss
these aspects more in depth later in Sec.~\ref{sec:trigv}.


\subsection{Numerical modelling}
\label{sec:numer}
We have seen in Sec.~\ref{sec:plane} that, for homogeneous pumps,
$\mathcal{F}_{f_p,\sigma_p} = f_p$, the conditions under which a
stable OPO switches on can be found analytically by assuming that
pump, signal, and idler states can be described by plane wave
fields~\eqref{eq:liopo} and therefore are characterised by single
wave-vectors $\vect{k}_{p,s,i}$ and by uniform currents, the intensity
and direction of which are given by
$\vect{k}_{p,s,i}$~\footnote{\ \label{footnote:curre} Given a complex
  field or wave-function, $|\psi(\vect{r},t)|e^{i\phi (\vect{r},t)}$,
  describing either a quantum particle of mass $m$ or a macroscopic
  number of particles condensed in the same quantum state, the current
  is defined as~\cite{pitaevskii03}:
  \begin{equation}
    \vect{j} (\vect{r},t) = \frac{\hbar}{m}
    |\psi(\vect{r},t)|^2 \nabla \phi (\vect{r},t) =
    |\psi(\vect{r},t)|^2 \vect{v}_s (\vect{r},t)\; ,
  \end{equation}
  following, with a slight abuse of notation, we will refer to the
  current as the gradient of the phase only, $\nabla \phi
  (\vect{r},t)$.}. However, for pumping lasers with a finite
excitation spot, $\mathcal{F}_{f_p,\sigma_p} (r)$, such as the ones
employed in experiments, one can only resort to a numerical
analysis~\cite{whittaker2005_b} of the coupled
equations~\eqref{eq:ppump}. A finite size pump implies that, in the
OPO regime, pump, signal, and idler states are broaden in momentum; as
a consequence, these states are going to be characterised by
non-trivial configurations of the currents (see
Fig.~\ref{fig:opocl}). We will see later on that these currents play a
crucial role in the occurrence and dynamics of both metastable and
spontaneous vortices in OPO.

\begin{figure}
\begin{center}
\includegraphics[width=1.0\linewidth,angle=0]{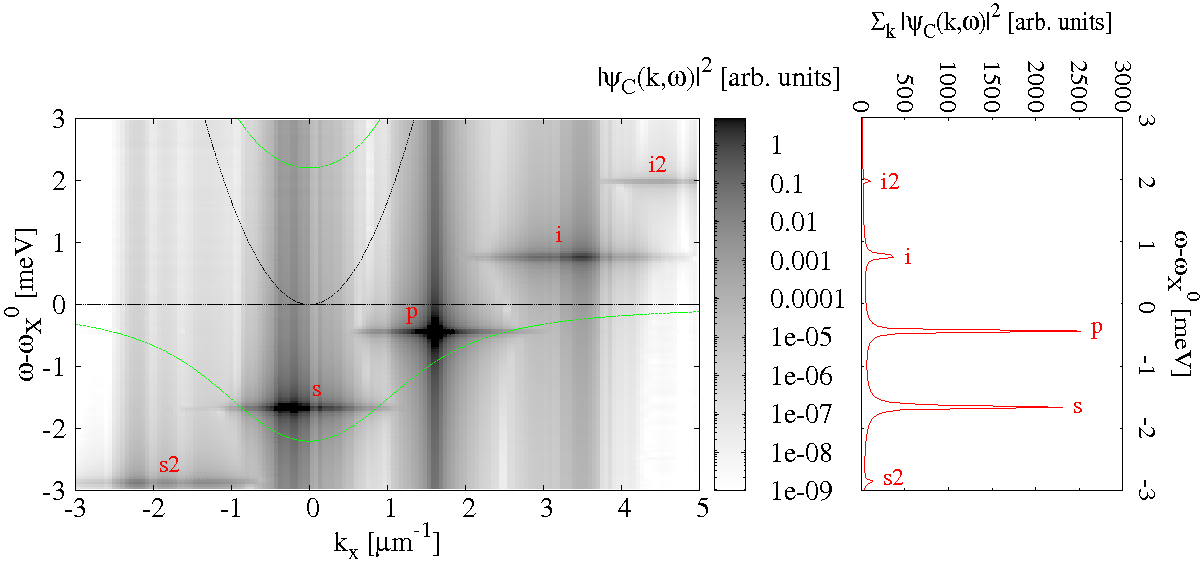}
\end{center}
\caption{Left panel: OPO spectrum $|\psi_{C,X} (\vect{k},\omega)|^2$
  for a top-hat pump of FWHM $\sigma_p=70$~$\mu$m and intensity
  $f_p=1.25 f_p^{\text{th}}$ above the threshold pump power for OPO,
  $f_p^{\text{(th)}}$. For this particular run we resonantly pump at
  $k_p=1.6$~$\mu$m$^{-1}$ in the $x$-direction, $\vect{k}_p =
  (k_p,0)$, and at $\omega_p-\omega_X^0=-0.44$~meV. Polaritons at the
  pump state undergo coherent stimulated scattering into the signal
  and idler states, which are blue-shifted with respect to the bare
  lower polariton (LP) dispersion (green dotted line) because of
  interactions. Cavity photon (C) and exciton (X) dispersions are
  plotted as gray dotted lines. Above threshold, as discussed in the
  text, we observe the population of the satellite states in addition
  to the one of signal and idler.  Right panel: Momentum integrated
  spectrum, $\sum_{\vect{k}} |\psi_{C,X} (\vect{k},\omega)|^2$, as a
  function of the rescaled energy $\omega - \omega_X^0$. Pump, signal,
  idler and satellite states are all equally spaced in energy by
  roughly $1.19$~meV.}
\label{fig:spect}
\end{figure}

\begin{figure}
\begin{center}
\includegraphics[width=1.0\linewidth,angle=0]{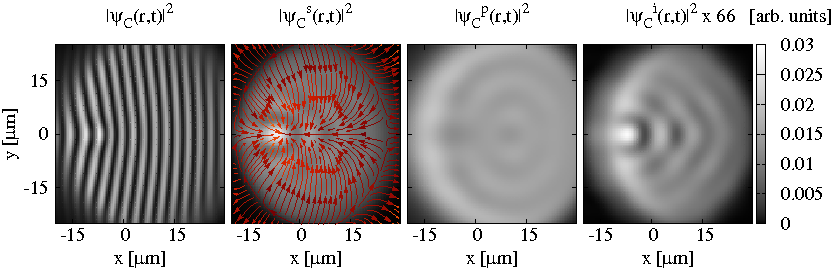}
\end{center}
\caption{Full emission (first panel) and filtered emission of signal
  (second), pump (third, with superimposed currents), and idler
  (fourth) states for the same parameters of Fig.~\ref{fig:spect}. We
  plot the rescaled currents of the signal in the second panel by
  subtracting the dominant uniform flow, i.e., $\nabla \phi_{C,X}^{s}
  - \vect{k}_{s}$.}
\label{fig:opocl}
\end{figure}

\begin{figure}
\begin{center}
\includegraphics[width=1.0\linewidth,angle=0]{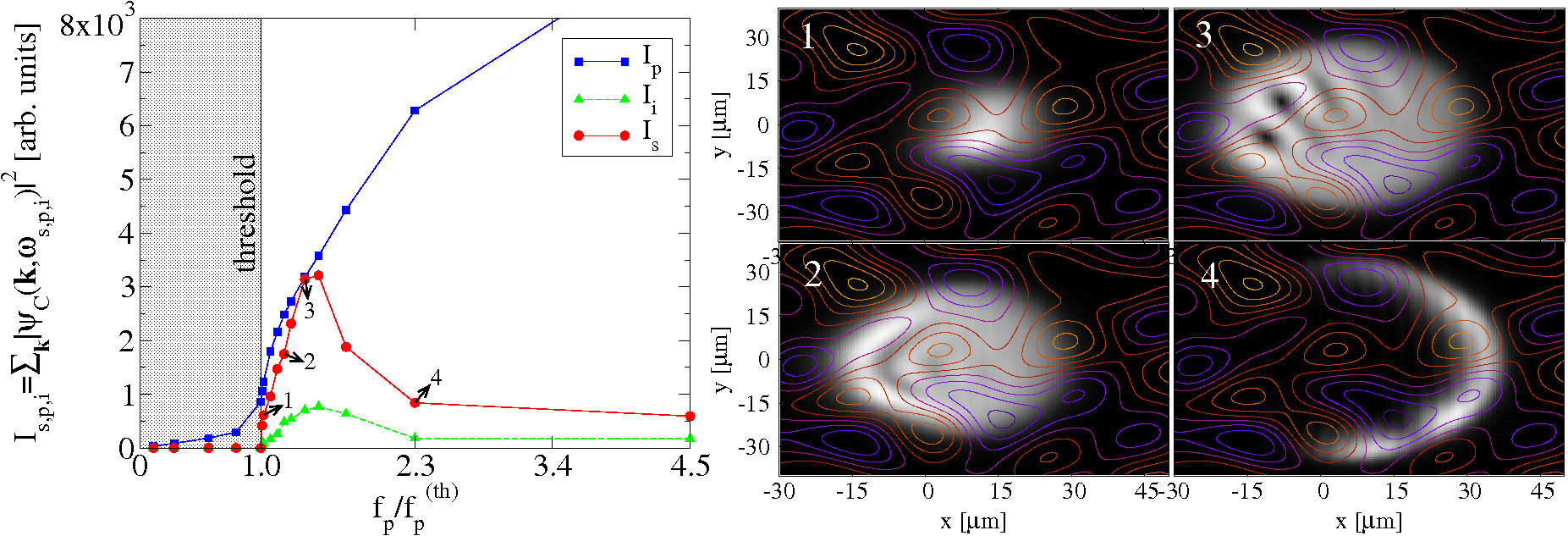}
\end{center}
\caption{Evolution of the signal, idler and pump state intensities as
  a function of the pump intensity $f_p/f_p^{\text{th}}$ (left). Space
  profiles of the filtered signal at different values of the pump
  intensity (right).  The parameters are the same as in
  Fig.~\ref{fig:spect}, with the addition of a photonic disorder
  potential $V_C(\vect{r})$, correlation length $\ell_d \simeq 20\
  \mu$m and strength $\sigma_d \simeq 0.1$~meV (contour-level lines in
  the panels on the right).}
\label{fig:thres}
\end{figure}

In particular, we numerically solve Eqs.~\eqref{eq:ppump} on a 2D grid
of typically $N\times N=2^8\times 2^8$ points and a separation of
$0.47$~$\mu$m (i.e., in a box $L\times L = 140$~$\mu$m$\times
140$~$\mu$m) by using a 5$^{\text{th}}$-order adaptive-step
Runge-Kutta algorithm. We have checked that our results are converged
with respect to both the resolution in space $L/N$ and the one in
momentum $\pi/L$. Note also that of course the extension of the
momentum box $k_{\text{max}}=\pi N/L $ has to be big enough to contain
the idler state. In the specific case of Figs.~\ref{fig:spect},
\ref{fig:opocl}, and~\ref{fig:thres}, we have chosen a smoothed
top-hat profile $\mathcal{F}_{f_p,\sigma_p} (r)$ with FWHM
$\sigma_p=70$~$\mu$m and (maximum) strength $f_p$ (later for
Fig.~\ref{fig:noise} we have chosen instead a FWHM
$\sigma_p=35$~$\mu$m). Considering the case of zero detuning,
$\delta=0$, we pump at $k_p=1.6$~$\mu$m$^{-1}$ in the $x$-direction,
$\vect{k}_p = (k_p,0)$, and at $\omega_p-\omega_X^0=-0.44$~meV,
i.e. roughly $0.5$~meV above the bare LP dispersion, and gradually
increase the pump strength until the OPO switches on. We find that
broader LP linewidths imply a wider range in pump strength of stable
OPO and for this reason we fix $\kappa_X=\kappa_C=0.26$~meV in these
particular runs.  We define $f_p^{\text{th}}$ as the pump strength
threshold for OPO emission --- here and in the following, we only
select OPO solutions which reach a dynamically stable steady state
(dynamical stability is studied in Sec.~\ref{sec:dynst}). In the case
of Figs.~\ref{fig:spect} and~\ref{fig:opocl}, the pump strength is
fixed just above threshold, $f_p=1.25 f_p^{\text{th}}$.

The numerical analysis provides the time evolution of both photon and
exciton fields either in space, $\psi_{C,X} (\vect{r},t)$, or in
momentum, $\psi_{C,X} (\vect{k},t)$. The OPO implies the simultaneous
presence of (at least) three states emitting at different momenta, and
therefore, at a fixed time $t$, the full emission $\psi_{C,X}
(\vect{r},t)$ is characterised by interference fringes. Because, as
for the pump, the dominant wave-vectors for signal and idler are in
the $x$-direction, the fringes are vertical, i.e. predominantly
oriented along the $y$-axis (see first panel of
Fig.~\ref{fig:opocl}). We plot the photon component only, which is
what can be measured experimentally. Note, however, that in cw
experiments emission is always integrated in time, which clearly
washes away the interference fringes. The OPO phase information can
instead be recovered by obtaining interference fringes with a
reference beam in a Michaelson configuration. In addition to the full
emission, either in space or momentum, one can also evaluate the
spectrum resolved in momentum $\psi_{C,X} (\vect{k},\omega)$ by taking
the Fourier transform in time of $\psi_{C,X} (\vect{k},t)$ (in
Fig~\ref{fig:spect}, a grid in time of $2^9$ points spaced by $0.3$~ps
has been used). As shown in Fig.~\ref{fig:spect}, for the chosen
parameters, a signal at $\omega_s-\omega_X^0=-1.66$~meV and an idler
at $\omega_s-\omega_X^0=0.75$~meV appear with a sharp $\delta$-like
emission in energy, which satisfies exactly the energy matching
condition~\eqref{eq:match}, $2\omega_p = \omega_s + \omega_i$, as
clearly shown by the momentum integrated spectrum on the right panel
of Fig.~\ref{fig:spect}. In contrast, the momentum distribution is
broad (because of the pump being finite size) and peaked respectively
at $k_s\simeq-0.2$~$\mu$m$^{-1}$ and $k_i\simeq 3.5$~$\mu$m$^{-1}$,
which only roughly satisfies the momentum matching condition, $2
\vect{k}_p = \vect{k}_s + \vect{k}_i$.
Note that the idler intensity is always weaker than the signal one
because of the small photonic component at the idler.
Further, note that, in addition to signal and idler, the spectrum also
shows the appearance of satellite states ($s_2, s_3, \dots$ and $i_2,
i_3, \dots$) all equally spaced at around $1.19$~meV one from the
other. As discussed at the end of Sec.~\ref{sec:plane}, their presence
is a consequence of the secondary parametric scattering processes $2s
\mapsto s_2 + p$, $2i \mapsto i_2 + p$, $2s_2 \mapsto s_2 + s_3$, and
so on, which trigger on automatically as soon as signal and idler have
finite populations. The occupation of the satellite states gets
gradually suppressed the further we move higher in energy above the
idler and lower in energy below the signal --- which is why they are
usually neglected in the plane wave approximation, as discussed in the
end of Sec.~\ref{sec:plane}. Note also that the satellite states just
described do not imply the presence of phase symmetries additional to
the $U(1)$ one described in Sec.~\ref{sec:golds}. These satellite
states therefore differ from the states which one could obtain as a
result of secondary instabilities, e.g. $2s \mapsto s_2' + s_2''$ with
$s_2''\ne p$ and $2i \mapsto i_2' + i_2''$ with $i_2''\ne p$, and
successive spontaneous symmetry breaking mechanism~\cite{cross93}.

%
In order to analyse the OPO properties, similarly to what is done in
experiments, it is also useful to filter the full emission in order to
select only the emission coming from the signal, pump or idler. This
can be equivalently done either filtering in momentum space in a cone
around the momenta $\vect{k}_{p,s,i}$ or filtering in energy, bringing
to the same results. We indicate the filtered spatial profiles of
pump, signal, and idler by
$|\psi_{C,X}^{p,s,i}(\vect{r},t)|e^{i\phi_{C,X}^{p,s,i}(\vect{r},t)}$.
The associated currents, $\nabla \phi_{C,X}^{p,s,i}$, are a
superposition of a dominant uniform flow $\vect{k}_{p,s,i}$ (which is
subtracted from the images of the second panel of
Fig.~\ref{fig:opocl}) and more complex currents (caused by the system
being finite size), which move particles from gain to loss dominated
regions.
Note that because we select only steady state OPO solutions, the
profiles of pump, signal, and idler,
$|\psi_{C,X}^{p,s,i}(\vect{r},t)|$, are time independent.
In addition, note that, the fact that the pump is shined on the
microcavity with a finite angle respect to the normal incidence,
implies that, for rotationally symmetric pump profiles, the symmetry
inversion $\vect{r} \mapsto -\vect{r}$ is broken in the direction of
the pump wave-vector $\vect{k}_p$. For example, if the pump is shined
on the $x$-direction, $\vect{k}_p = (k_p,0)$, as in the case of
Fig.~\ref{fig:opocl}, only the symmetry $y \mapsto -y$ is left
intact. Clearly, this symmetry, while allowing vortex-antivortex
pairs, does not in principle permit OPO solutions carrying single
vortices, which can spontaneously appear in presence of a symmetry
breaking perturbation, such as disorder (next paragraph) or a noise
pulse (see Sec.~\ref{sec:dynst}).

The typical changes of the signal space profile as the pump power is
increased above threshold, together with the pump, signal, and idler
intensities, are shown in Fig.~\ref{fig:thres}. For these runs we fix
the parameters, such as $\omega_p$, $\vect{k}_p$, and the pumping spot
size $\sigma_p$, as in Fig.~\ref{fig:spect}, but we also include a
static photonic disorder potential --- see Eq.~\eqref{eq:polai}. In
particular, here we consider a disorder potential with zero average,
$\langle V_C(\vect{r}) \rangle =0$ and a spatial distribution,
\begin{equation}
  \langle V_C(\vect{r}) V_C(\vect{r}') \rangle = \sigma_d^2
  e^{-|\vect{r}-\vect{r}'|^2/2 \ell_d^2}\; ,
\end{equation}
with a correlation length $\ell_d \simeq 20\ \mu$m and strength
$\sigma_d \simeq 0.1$~meV. Below threshold the system is in a
pump-only state. By increasing the pump power $f_p$, above threshold,
the OPO signal first switches on only in a small (compared with the
pump spot FWHM $\sigma_p=70\ \mu$m) region (see inset 1). At $f_p =
1.2f_p^{\text{th}}$ (inset 2) the signal becomes large and quite
homogeneous, though, already at $f_p = 2.3f_p^{\text{th}}$, the OPO
signal starts switching off in the middle (inset 4), and then it
slowly switches off everywhere. A similar behaviour has been found in
the numerical simulations of Ref.~\cite{whittaker2005_b} (though there
a small pump beam of FWHM $\sigma_p\sim20\ \mu$m has been used), as
well as observed experimentally in Ref.~\cite{sanvitto06}.

The qualitative behaviour of the signal (as well as the idler)
profiles, in particular their switching on and then off, as a function
of the pump power that we have just described for a disordered sample
is very similar to the case of an OPO in a homogeneous
sample~\footnote{\ Note also that we find that the value of the pump
  threshold for OPO is not altered by the presence of a weak photonic
  disorder.} (i.e., with no photonic disorder, $V_C(\vect{r})=0$).
One of the main differences is that for homogeneous samples the
profiles are $y \mapsto -y$ symmetric, while this symmetry is
explicitly broken by the photonic disorder. In addition, the
fundamental difference between the homogeneous and the disordered
case, is that the presence of photonic disorder promotes stable vortex
solutions in large pump spot OPOs at intermediate pumping strengths,
$f_p \simeq 1.4f_p^{\text{th}}$ --- such as the one shown in panel 3
of Fig.~\ref{fig:thres} which carries two vortices.
Single or multiple vortex solutions are generally not allowed in the
homogeneous case because of the $y \mapsto -y$ symmetry, which instead
only allows pairs of vortex-antivortex solution $y \mapsto -y$
symmetric. In large pump spots, such as the one of
Fig.~\ref{fig:thres}, vortex-antivortex solutions in the clean case
tend to be dynamically unstable, i.e. easily destabilised by a weak
noise pulse, while, as analysed later in Sec.~\ref{sec:stabv},
spontaneous vortex solutions in homogeneous cavities can be stabilised
by a small pump spot (see Fig.~\ref{fig:noise}) confining the vortex
inside. Note finally that, spontaneous vortices in disordered cavities
with a large pump spots are not pinned into minima of the disorder
potential, rather, as analysed in the Sec.~\ref{sec:aver}, are the OPO
steady state currents in the signal to play an essential role in the
stabilisation of vortices.



\subsection{Vortex phase and profile}
\label{sec:vorte}
\begin{figure}
\begin{center}
\includegraphics[width=0.9\linewidth,angle=0]{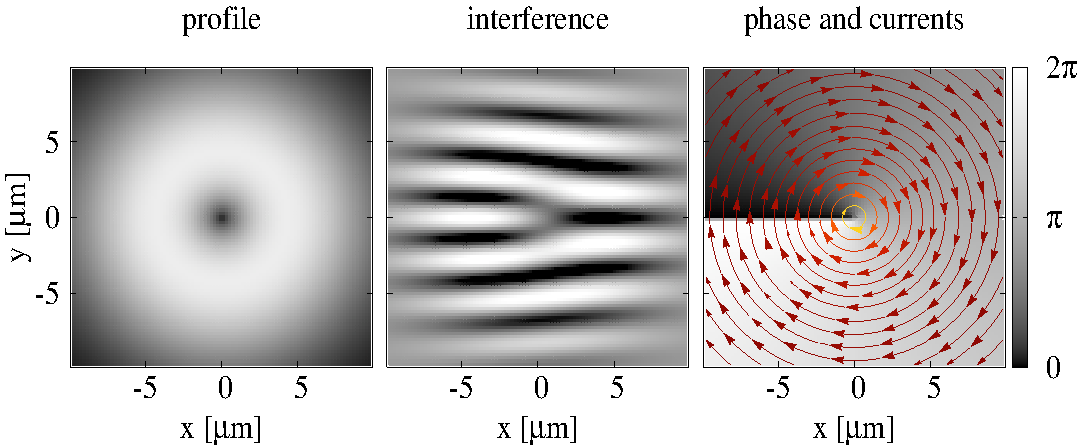}
\end{center}
\caption{Typical profile (left), phase and currents~\eqref{eq:curre}
  (right), and interference fringes (middle) of an $m=-1$
  vortex~\eqref{eq:vorte}.}
\label{fig:examp}
\end{figure}

Before moving on to describe the occurrence of stable vortices in OPO,
and, later, the onset and dynamics of metastable vortices, let us
briefly remind the definition of a quantised vortex in an
\emph{irrotational} fluid. In general, a quantised vortex with charge
$m$ is described by a wave-function,
\begin{equation}
  \psi(\vect{r}) = \psi_0 (r) e^{i m \varphi (\vect{r})} \; ,
\label{eq:vorte}
\end{equation}
the phase of which, $m \varphi (\vect{r})$, linearly winds around the
vortex core from $0$ to $2\pi m$ (with $m$ integer) --- i.e., in
cylindrical coordinates centered at the vortex core, $\varphi$ is the
azimuthal angle. This implies that the vortex carries a quantised
angular momentum, $\hbar m$. In addition, the phase has a branch-cut
and therefore is not defined at the vortex core, implying the vortex
wave-function has to be zero at the vortex core.
An example of an $m=-1$ vortex, with $\psi_0 (r) = r e^{-
  r^2/(2\sigma^2_{v})}$, has been plotted on the left panel of
Fig.~\ref{fig:examp}. Here, the phase winds clock-wise around the
core, from $0$ to $2\pi$, and therefore the vortex current,
\begin{equation}
  \nabla \varphi (\vect{r}) = \frac{\hat{\varphi}}{r} \; , 
\label{eq:curre}
\end{equation}
is constant at fixed distances from the vortex core, $r$, while
decreases inversely proportional to the distance (right panel of
Fig.~\ref{fig:examp}). 
Contrast this with the case of a \emph{rotational} vortex in a
classical fluid which rotates as a solid body with an angular velocity
$\Omega$: Now, the fluid tangential velocity is zero at the vortex
core and increases linearly with the distance, i.e., $\vect{v}_\varphi
= \Omega r \hat{\varphi}$.
Quantised vortices can be detected in interference fringes (middle
panel of Fig.~\ref{fig:examp}) as fork-like dislocations, the
difference in arms giving the charge $|m|$ of the vortex.
%

\subsection{Stable vortices in a small sized OPO}
\label{sec:stabv}
As explained later in Sec.~\ref{sec:trigv}, spontaneous stable
vortices differ from metastable vortices (described in
Sec.~\ref{sec:metas}): Metastable vortices can only be injected
externally, e.g. by an additional Laguerre-Gauss beam probe, into an
otherwise stable symmetric state, and their persistence is due to the
OPO superfluid properties~\cite{sanvitto10,marchetti10}. The
metastable vortex is a possible but not unique stable configuration of
the system.
In contrast, as for non-resonantly pumped
polaritons~\cite{keeling08,lagoudakis08}, the appearance of
spontaneous vortices is not a consequence of the polariton condensate
being superfluid, but rather to the presence of currents related to
the non-equilibrium nature of these condensates. This strongly differs
from the case of equilibrium superfluids, the ground state of which is
flow-less. Later, in Sec.~\ref{sec:other} we will briefly discuss how,
for polaritons non-resonantly injected into a microcavity, the
presence of a confining potential can generate currents favourable to
the spontaneous formation of vortices~\cite{lagoudakis08,nardin10} and
vortex lattices~\cite{keeling08}.

For resonant excitation, currents arise in the OPO regime due to the
simultaneous presence of pump, signal, and idler emitting at different
momenta, as well as by the fact the system is finite size (see
Fig.~\ref{fig:spect}). We have seen in Fig.~\ref{fig:thres} that,
similarly to non-resonantly pumped polaritons, the presence of a
disorder potential can lead to the spontaneous appearance of
vortices. However, it is remarkable that, even in the absence of
disorder or trapping potentials, the OPO system can undergo
spontaneous breaking of the $y \mapsto -y $ symmetry and become
unstable towards the formation of a quantised vortex state with charge
$m=\pm 1$ if the size of the OPO is small
enough~\cite{marchetti10}. This is the subject of this section.
Further, as discussed in some detail later in Sec.~\ref{sec:heali},
like for equilibrium superfluids, both stable and metastable vortices
are characterised by a healing length which is determined by the
parameters of the OPO system alone.
Spontaneous stable vortex solutions are robust to noise
(Sec.~\ref{sec:dynst}) and to any other external perturbation, and
thus should be experimentally observable.  However, while spontaneous
vortex solutions in OPO have been observed for a toroidal pump
spot~\footnote{\label{private}\ D. Sarkar (University of Sheffield),
  private communication.}, so far they have not been observed in OPO
with a `simply connected' pump profile, e.g., either a Gaussian or a
top-hat.

\begin{figure}
\begin{center}
\includegraphics[width=0.9\linewidth,angle=0]{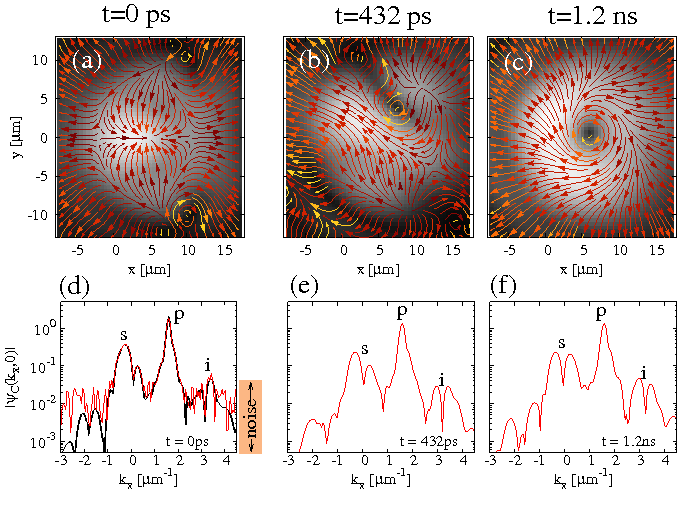}
\end{center}
\caption{Appearance of a spontaneous stable vortex solution in a
  homogeneous small sized OPO. Filtered signal profile
  $|\psi_C^{s}(\vect{r},t)|$ with superimposed currents $\nabla
  \phi_C^{s} (\vect{r},t)$ (upper panels (a-c)) and full momentum
  emission $|\psi_C(k_x,0,t)|$ (lower panels (d-f), in arb. units) at
  three different times: $t=0$ (a,d), $t=432$ps (b,e), and $1.2$ns
  (c,f). At $t=0$ a pulsed weak random noise of strength $0.01$ (see
  text) is added to the OPO steady state (in (d) both OPO momentum
  profiles without and with the added noise are shown for comparison)
  and at $t=432$ps a vortex, with $m=-1$, enters the signal and
  settles into a steady state. Note that, because of phase matching
  conditions, the presence of an $m=-1$ vortex in a signal implies the
  presence of an $m=1$ antivortex in the idler. A vortex (antivortex)
  in the signal (idler) space emission appears also as a dip in
  momentum space at the signal (idler) momentum (e,f). Parameters
  used: smoothed top-hat pump with FWHM $\sigma_p=35\mu$m, pump
  strength $f_p=1.12 f_p^{\text{(th)}}$, $k_p=1.6$~$\mu$m$^{-1}$ in
  the $x$-direction, $\omega_p-\omega_X^0=-0.44$~meV, zero detuning
  $\delta=0$, and $\kappa_X=\kappa_C=0.22$~meV. Adapted
  from~\cite{marchetti10}.}
\label{fig:noise}
\end{figure}
%
\subsubsection{Dynamical stability}
\label{sec:dynst}
As mentioned in Sec.~\ref{sec:numer}, if the pump is shined on the
$x$-direction, $\vect{k}_p = (k_p,0)$, only the symmetry $y \mapsto
-y$ is left intact in the system. Clearly, this symmetry, allows for
OPO solutions where the signal (and therefore also the idler) have
vortex-antivortex pairs, with the vortex core position at $(x_c, y_c)$
and the antivortex core position at $(x_c, -y_c)$. However, both
single and multiple vortex solutions explicitly break the $y \mapsto
-y$ symmetry and cannot be accessed by the dynamics --- Note that two
vortices located at opposite sides with respect to the $x$-axis break
the $y \mapsto -y$ symmetry because of the currents.

In order to check the dynamical stability of OPO states, one has to
add small fluctuations to the steady state mean-field solution: The
existence of modes with positive imaginary part in the excitation
spectrum indicate dynamical instability towards the growth of
different modes. The dynamical stability analysis for OPO described in
within the plane-wave approximation of Sec.~\ref{sec:plane} has been
discussed in Refs.~\cite{whittaker05,wouters06b}.
Equivalently, stability can be numerically checked by introducing a
weak noise. In particular, we add white noise as a quick
($\delta$-like in time) pulse at a certain time $t_0$ to both modulus
and phase of excitonic and photonic wavefunctions in momentum space,
$|\psi_{X,C}(\vect{k},t)| e^{i \phi_{X,C}(\vect{k},t)}$:
\begin{align*}
  |\psi_{X,C}(\vect{k},t_0)| &\mapsto |\psi_{X,C}(\vect{k},t_0)| +
  \delta |\psi_{X,C}(\vect{k})| \\ \phi_{X,C}(\vect{k},t_0) &\mapsto
  |\phi_{X,C}(\vect{k},t_0) + \delta \phi_{X,C}(\vect{k}) \; .
\end{align*}
Both $\delta |\psi_{X,C}(\vect{k})|$ and $\delta \phi_{X,C}(\vect{k})$
are white noise functions, with an amplitude $2\pi$ for the the phase
$\delta \phi_{X,C}(\vect{k})$, while the amplitude of the noise in the
modulus $\delta |\psi_{X,C}(\vect{k})|$ is specified in units of the
maximum of the pump intensity in momentum space.

\begin{figure}
\begin{center}
\includegraphics[width=0.8\linewidth,angle=0]{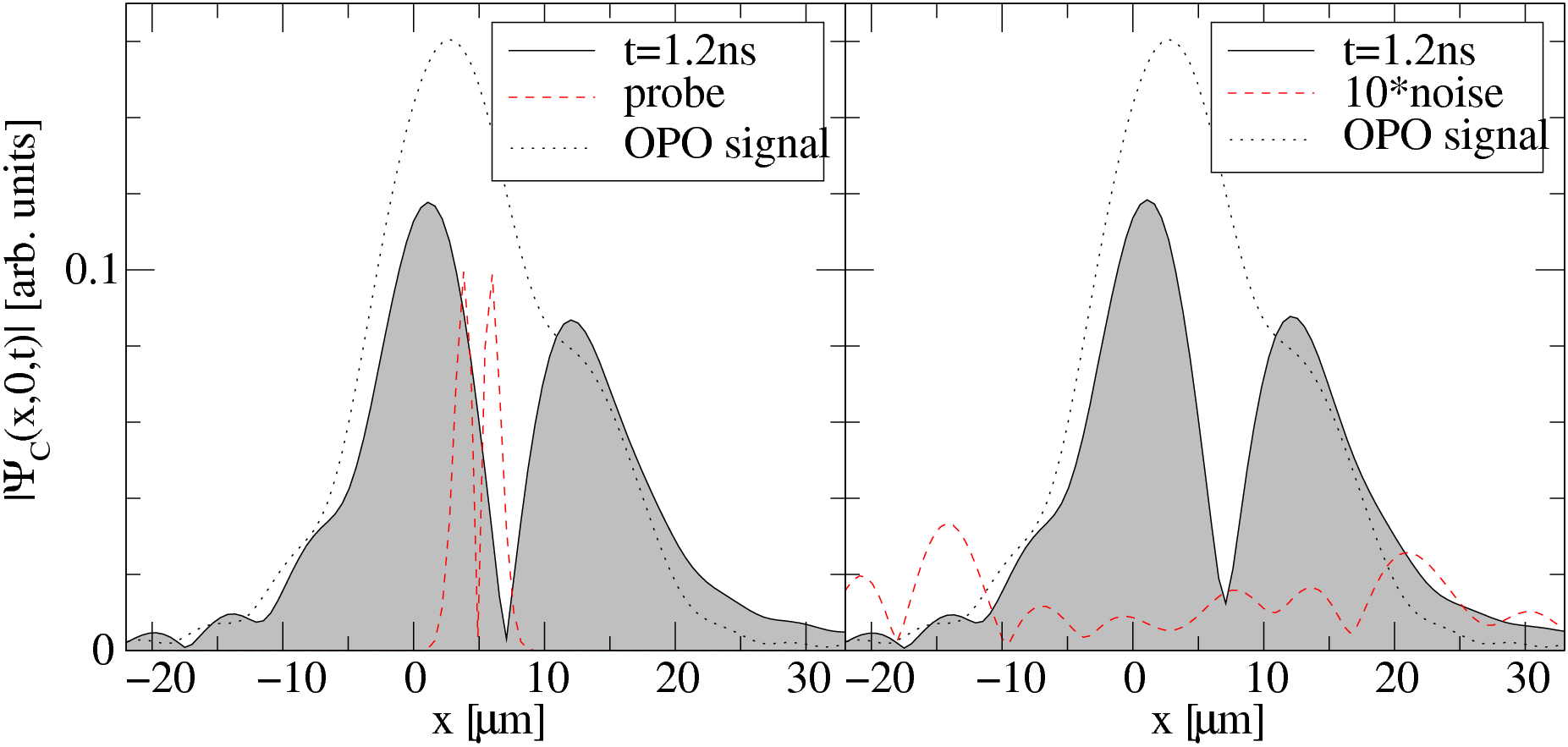}
\end{center}
\caption{Steady state filtered signal profile (dotted line) $\psi_C^s
  (x,0,t)$ for $y\simeq 0$ before the arrival of either a
  Laguerre-Gauss vortex probe~\eqref{eq:probe} with $\sigma_{pb}
  \simeq 1\mu$m (left panel, red dashed line) or a noise pulse of
  strength 0.01 (right panel, red dashed line) --- same OPO conditions
  as Fig.~\ref{fig:noise}. After the arrival of any perturbation
  breaking the $y\mapsto -y$ symmetry, the same vortex with charge
  $m=\pm 1$ (solid shaded curve) stabilises into the
  signal. [From~\cite{marchetti10} {\tt ask for copyright
      permission!}]}
\label{fig:profi}
\end{figure}

Following this procedure, we have been able to single out symmetric
OPO states, as shown in Fig.~\ref{fig:noise}(a), which are unstable
towards the spontaneous formation of stable vortex solutions. After
the $y \mapsto -y$ symmetry is broken by the noise pulse, we have
observed a vortex with quantised charge $m=\pm 1$ ($m=\mp 1$) entering
and stabilising into the OPO signal (idler) --- Note that parametric
scattering constrains the phases of pump, signal, and idler by
$2\phi_p=\phi_s+\phi_i$ (see Sec,~\ref{sec:golds}), therefore an
$m=-1$ vortex in the signal at a given position implies an $m=1$
antivortex in the idler at the same position and vice versa.
In the case of Fig.~\ref{fig:noise} and the right panel of
Fig.~\ref{fig:profi}, the noise strength is $0.01$ and $432$ps after
the noise pulse, a vortex with $m=-1$ ($m=+1$) enters the signal
(idler) and stabilises. The strength of the noise has no relevance on
the final steady state, and in particular it can be infinitesimally
weak. Different noise strengths do only affect the \emph{transient}
time the system needs to accommodate the vortex and reach the new
steady configuration. 
We have in addition examined whether the vortex steady state is
dynamically stable by applying an additional noise pulse. For weak
noise, with a strength up to $0.1$, the vortex is stable and can only
drift around a little before settling again into the same state. For
strong noise, with strength $1$ and above, the vortex gets washed
away, but after a transient period, the very same state enters and
stabilises again into the signal, with the possibility of flipping
vorticity~\footnote{\ When generated by a noise pulse, both stable and
  metastable vortices have equal probability to have either charge
  $\pm1$. Similarly, when vortices are triggered via a Laguerre-Gauss
  probe, their vorticity can flip during the transient period. In
  particular, flipping can follow the appearance of two antivortices
  at the edge of the signal, one recombining with the triggered
  vortex. Note that the vorticity flipping conserves the total orbital
  angular momentum, in the sense that when for the signal $m$ flips,
  say, from $+1$ to $-1$, for the idler the opposite happens, i.e. $m$
  flips from $-1$ to $+1$.}. Different noise strengths do not affect
the final steady state, but only the transient time.

As discussed later in Sec.~\ref{sec:trigv}, one can alternatively
break the $y \mapsto -y$ symmetry by a pulsed vortex
probe~\eqref{eq:probe}, and assess whether the stable steady state is
in any way dependent on the external perturbation. The homogeneous OPO
states which are unstable towards the spontaneous formation of stable
vortices following a white noise pulse, exhibit the same instability
following a vortex Laguerre-Gauss (LG) probe pulse (see the left panel
of Fig.~\ref{fig:profi}). The steady state vortex is independent on
both the probe intensity $f_{pb}$ and size $\sigma_{pb}$, however the
weaker the probe the longer the vortex takes to stabilise, between
$30$ and $400$ps for our system parameters. As shown in
Fig.~\ref{fig:profi}, the stable vortex following the LG probe is
exactly the same as the one triggered by a weak white noise,
indicating that the probe acts only as a symmetry breaking
perturbation.

Summarising, one can find OPO conditions where the $y \mapsto -y$
symmetric solution is dynamically unstable and any symmetry breaking
perturbation allows the signal and idler to relax into a stable steady
state carrying a vortex with charge $\pm 1$.  For homogeneous
cavities, i.e., in absence of any disorder or confining potential, we
found that this requires either a small Gaussian or small top-hat like
pump spot which can confine the vortex inside or a doughnut-shape pump
spot. Instability of the uniform state to spontaneous pattern (e.g.,
vortex) formation is a typical feature of systems driven away from
equilibrium~\cite{cross93}. Similarly we find conditions for which the
uniform OPO solution is unstable to spontaneous formation of a
quantised vortex. In alternative, a disorder potential breaks the
symmetry explicitly and allows the pinning of stable vortex solutions
in OPO, which is less surprising.

\subsubsection{Healing length}
\label{sec:heali}
In contrast to their classical counterpart, quantised vortices with
the same angular momentum $|m|$ are all identical, with a size (or
healing length) determined by the system non-linear
properties~\cite{pitaevskii03}. In the case of a superfluid in
equilibrium with a typical interaction energy $gn$ ($n$ is the average
density) and mass $m$, the healing length, $\xi=1/\sqrt{2mg n}$, is
the typical distance over which the condensate wave-function recovers
its `bulk' value around a perturbation. In particular, for an $|m|=1$
vortex~\eqref{eq:vorte}, $\xi$ is the typical size of the vortex.

Similarly, in OPO, one case show that, like in equilibrium
superfluids, both stable (see Sec.~\ref{sec:stabv}) and metastable
(see Sec.~\ref{sec:trigv}) vortices are characterised by a healing
length which is determined by the parameters of the OPO system
alone. In particular, shape and size of the metastable vortices
described in Sec.~\ref{sec:trigv} are independent on the external
probe.  In the case of vortices in OPO, an approximate analytical
expression for the vortex healing length can be derived for
homogeneous pumping~\cite{marchetti10,krizhanovskii10}, assuming that
only signal and idler can carry angular momentum with opposite sign,
$\pm m$, $\psi^{s,i} (\vect{r}) = \sqrt{n_{s,i}} e^{i \vect{k}_{s,i}
  \cdot \vect{r}} e^{\pm i m \varphi} \Psi^{s,i} (r)$, while the pump
remains in a plane-wave state, $\psi^{p} (\vect{r}) = \sqrt{n_{p}}
e^{i \vect{k}_{p} \cdot \vect{r}}$, as also supported by our numerical
analysis. For pump powers close to OPO threshold, it can be
shown~\cite{marchetti10,krizhanovskii10} that signal and idler steady
state spatial profiles are locked together and satisfy the following
complex GP equation
\begin{equation*}
  \left[-\Frac{1}{2m_C} \left(\frac{d^2}{dr^2} + \frac{1}{r}
    \frac{d}{dr} - \frac{m^2}{r^2}\right) + \alpha \left({\Psi^{s}}^2
    -1 \right)\right]\Psi^{s} = 0 \; ,
\end{equation*}
where $|\alpha| \simeq g_X \sqrt{n_s n_i}$. 
This equation describes a vortex profile~\cite{pitaevskii03} with a
healing length given by:
\begin{equation}
  \xi = \left(2 m_C g_X \sqrt{n_s n_i}\right)^{-1/2}\; .
\label{eq:estim}
\end{equation}
This expression is similar to the one of an equilibrium superfluid,
with the condensate density replaced by the geometric average of
signal and idler densities. Further above threshold, one can show that
signal and idler profiles are no longer locked together, and that they
start to develop different radii. In both the simulations of
Figs.~\ref{fig:profi} and~\ref{fig:thres}, we find $\xi \simeq 4\mu$m,
compatible with the estimate~\eqref{eq:estim}. 



In Ref.~\cite{krizhanovskii10}, vortices in OPO have been created in a
controlled manner by adding a weak continuous probe in resonance to
the signal. Even if the phase freedom of the OPO system is explicitly
broken in this configuration by the vortex cw probe, because the ratio
of the probe to signal power density is low, the size of the vortex
has been demonstrated to be determined by the OPO non-linear
properties only rather than by the imprinting probe. In particular, a
systematic study of the decrease of the vortex core radius with
increasing pump power above threshold has allowed to confirm the
behaviour described by the Eq.~\eqref{eq:estim}.



\section{Triggered optical parametric oscillator regime}
\label{sec:topor}
Before moving on into the description of metastable vortices in OPO,
i.e. vortices which are transferred by a pulsed vortex probe into the
OPO signal and idler, and their relation to superfluidity
(Sec.~\ref{sec:trigv}), we describe here first the effect of an
additional pulsed probe on OPO in general terms.
As described previously, in the OPO regime, polaritons are
continuously injected into the pump state, and undergo coherent
stimulated scattering into the signal and idler states. The OPO is a
steady state regime, where the filtered profiles of signal, idler, and
pump, $|\psi_{C,X}^{p,s,i}(\vect{r},t)|$, are time independent. This
also is reflected in the typical flat dispersion around pump, signal,
and idler which can be observed in the OPO spectra (see
Fig.~\ref{fig:spect}). The \emph{group velocity} of pump, signal, and
idler, defined as the derivative of the energy dispersion at
$\vect{k}_{p,s,i}$, is therefore zero. This however does not mean that
there is no flow of polaritons, which instead is described by the
\emph{phase velocity} or current, $\nabla \phi_{C,X}^{p,s,i}$ (see
footnote~\ref{footnote:curre} on page~\pageref{footnote:curre}), with
a dominant uniform flow given approximatively by $\vect{k}_{p,s,i}$.

In resonantly pumped polaritons, in order to initiate a travelling
wave-packet characterised by a finite group velocity, one needs to use
an additional pulsed laser beam on top of the cw one. The description
of the system is therefore still in terms of the
Eqs.~\eqref{eq:ppump}, with a total pump term given by the sum of the
cw laser~\eqref{eq:pumpe} and a probe beam $F_{pb}(\vect{r},t)$:
\begin{equation}
  F(\vect{r},t) = F_{p}(\vect{r},t) + F_{pb}(\vect{r},t) \; .
\label{eq:pumpr}
\end{equation}
For the moment being we will consider the generic case of a pulsed
probe with a Gaussian space profile, shined at a momentum and energy
$\{\vect{k}_{pb},\omega_{pb}\}$~\footnote{\ Note that, differently
  from the cw laser beam, the energy distribution spectrum of which is
  essentially $\delta$-like, a pulsed beam has an intrinsic width in
  energy, proportional to the inverse pulse duration,
  $\sigma_t^{-1}$.}:
\begin{equation}
  F_{pb}(\vect{r},t) = f_{pb}
  e^{-|\vect{r}-\vect{r}_{pb}|^2/(2\sigma^2_{pb})} e^{i (\vect{k}_{pb}
    \cdot \vect{r} - \omega_{pb} t)} e^{-(t-t_{pb})^2/(2\sigma^2_{t})}
  \; .
\label{eq:probe}
\end{equation}
A pulse duration of $3$~ps (defined as the FWHM in time of
$F_{pb}(\vect{r},t)$) corresponds to $\sigma_t=1.3$~ps. The idea,
first introduced by Ref.~\cite{amo09}, is that the pulsed probe
triggers parametric scattering~\footnote{\ If the cw pump drives the
  system into the OPO regime, then the parametric scattering triggered
  by the pulsed probe will be in addition to the one related to
  OPO. However, as discussed later, the TOPO regime can be reached
  also in absence of the OPO.} between the probe state at momentum and
energy $\{\vect{k}_{pb},\omega_{pb}\}$ and a \emph{conjugate} state at
$\{\vect{k}_{c} = 2\vect{k}_p-\vect{k}_{pb} , \omega_{c} = 2\omega_p -
\omega_{pb}\}$ --- because one can either have $\vect{k}_{pb} >
\vect{k}_{p}$ or $\vect{k}_{pb} < \vect{k}_{p}$, we use the state
labels `probe' and `conjugate', rather than `signal' and `idler'; by
doing so, one also doesn't confuse the states generated by the OPO
with the additional ones generated by the probe. Both probe and
conjugate states are travelling decaying states which can evolve
freely from the laser probe constraints once the pulse switches
off. Such states are referred to as triggered-OPO (TOPO) states. Note
that a TOPO can be triggered in two regimes: either (i) in a regime
where the cw laser drives the system above threshold for OPO, in which
case the probe and conjugate states are the extra population states on
top of the steady-state OPO signal and idler states, or (ii) when no
OPO is present, i.e. for the cw pump strength below threshold. For
simplicity, the numerical analysis discussed below in
Sec.~\ref{sec:thtop} is conducted in the regime (ii), but we have
checked that the qualitative results also hold in the regime (i) ---
where, now, the steady state OPO population needs to be subtracted so
that one studies the properties of the population triggered by the
probe only.

\begin{figure}
\begin{center}
\includegraphics[width=1.0\linewidth,angle=0]{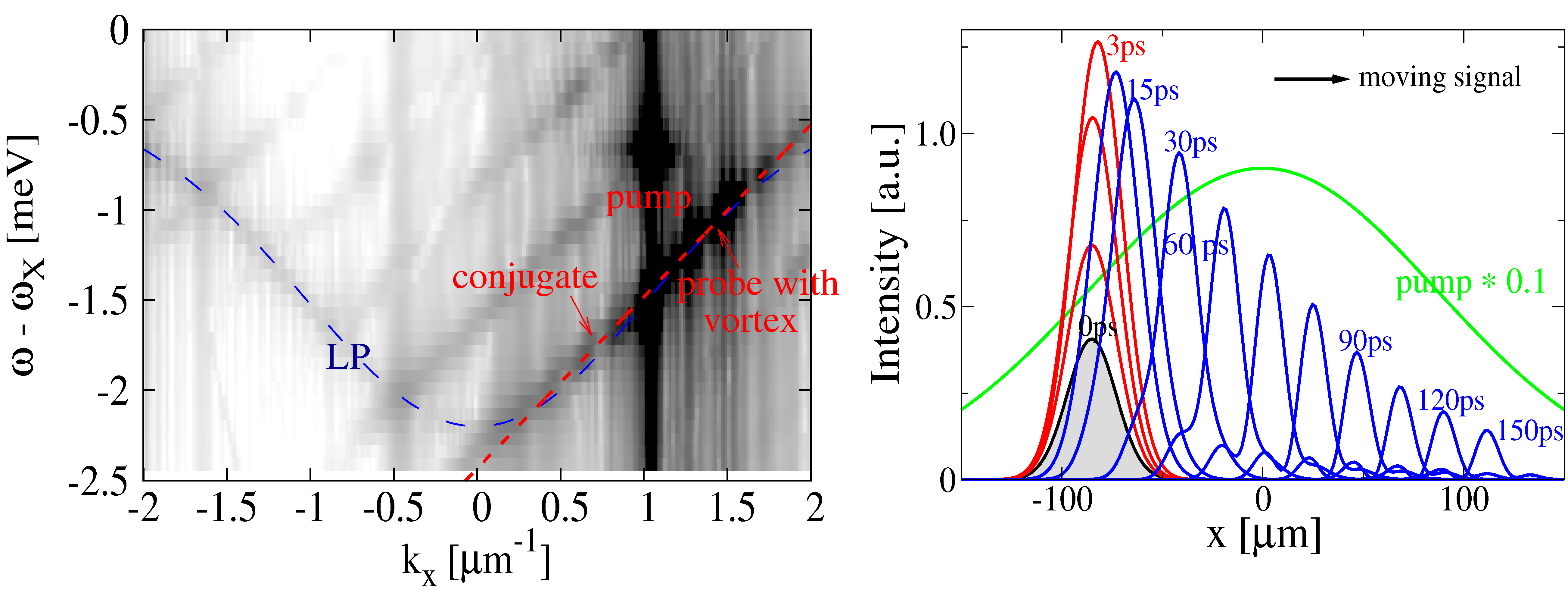}
\end{center}
\caption{Spectrum (left) and spatial profiles of pump and filtered
  signal $|\psi_C^{s}(x,0,t)|$ (right) for the TOPO regime. A short,
  $\sigma_t=1$ps, $m=2$ Laguerre-Gauss~\eqref{eq:vprob} (left) or
  Gaussian $m=0$~\eqref{eq:probe} (right) probe shined at
  $\vect{k}_{pb} = (1.4,0)~\mu\text{m}^{-1}$ triggers the propagating
  probe and conjugate states, which lock to the same group velocity
  (for these simulations we fix $\kappa_{X}=0$ and
  $\kappa_{C}=0.02$meV). Adapted from~\cite{szymanska10}. }
\label{fig:bulets}
\end{figure}
\begin{figure}
\begin{center}
\includegraphics[width=0.5\linewidth,angle=0]{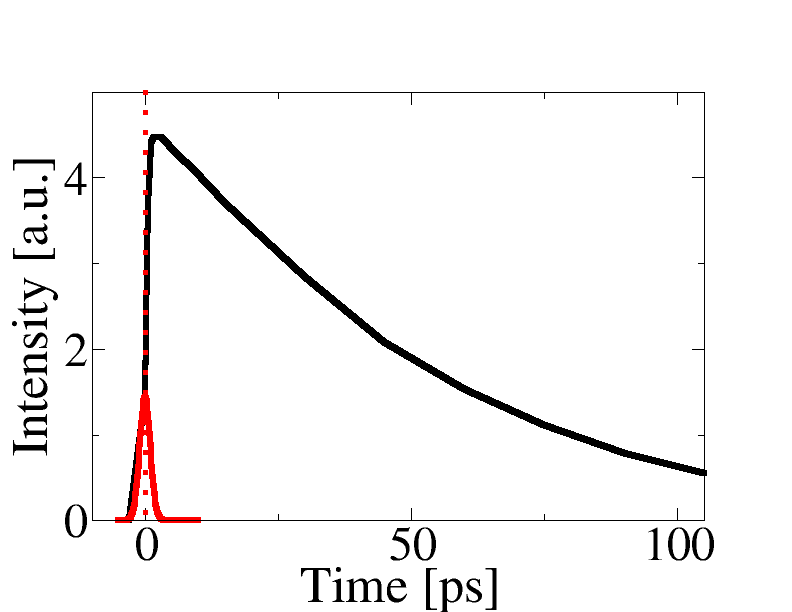}
\end{center}
\caption{TOPO signal intensity (black line) and the intensity of an
  external Gaussian probe (red line) as a function of time. Parameters
  are the same ones of the right panel in Fig.~\ref{fig:bulets}. The
  TOPO signal gets initially strongly amplified, then decays slowly,
  and finally exponentially.}
\label{fig:total_int}
\end{figure}
%
\subsection{Theoretical description of the TOPO}
\label{sec:thtop}
{\tt cross-refer to Fabrice Laussy' chapter}
In order to analyse the dynamical evolution of a TOPO wave-packet, we
study numerically the time-dependent solutions of the
equations~\eqref{eq:ppump}, with a total pump given
by~\eqref{eq:pumpr} and~\eqref{eq:probe}.
The probe triggers parametric scattering between a probe state and a
conjugate state. In the majority of cases, as discussed in
Ref.~\cite{szymanska10}, the parametric scattering is too weak to
induce any significant amplification, and an exponential decay of both
probe and conjugate populations is observed immediately after the
probe $F_{pb}(\vect{r},t)$ switches off. Here, the spectrum shows a
strong emission from the pump state and a weak emission from the LP
states mainly at momenta $\vect{k}_{pb}$ and $\vect{k}_c$.

\begin{figure}
\begin{center}
\includegraphics[width=1.0\linewidth,angle=0]{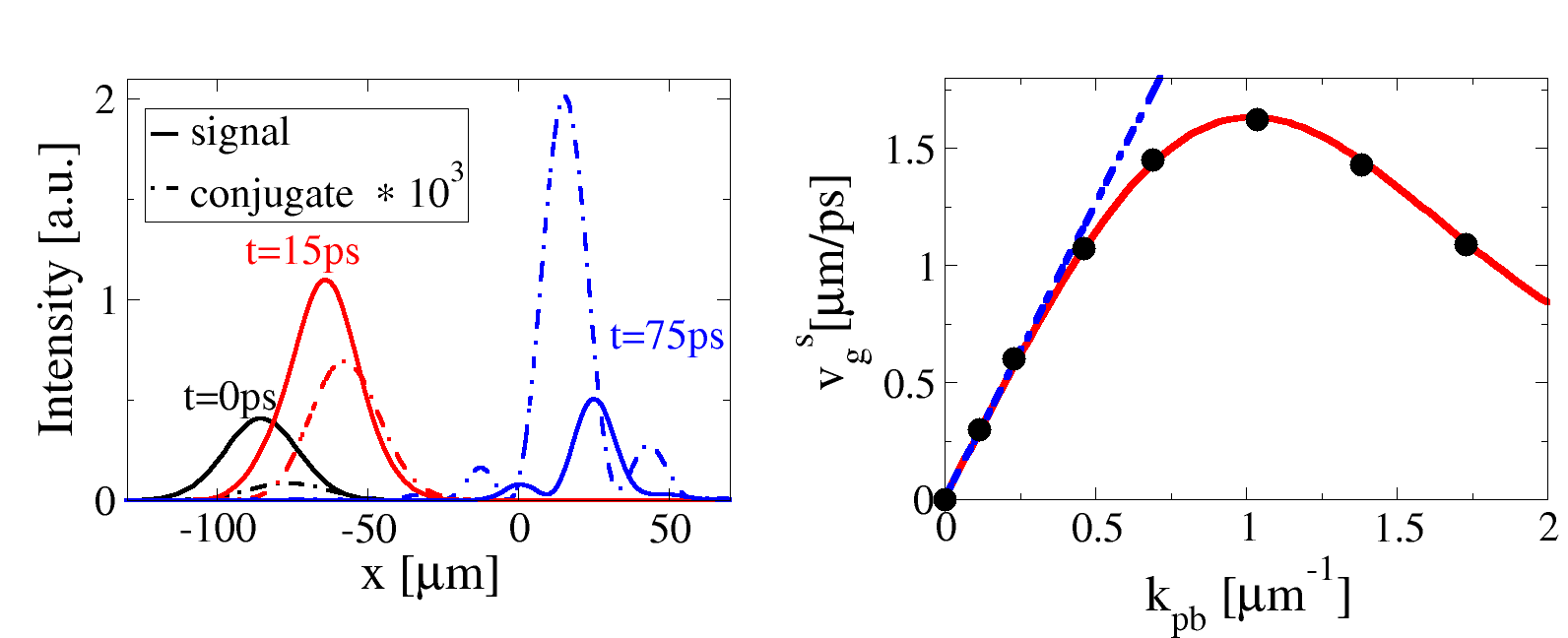}
\end{center}
\caption{Left panel: Probe (solid lines) and conjugate (dashed lines)
  density profiles at different times after the arrival of the probe
  (same parameters as in Fig.~\ref{fig:bulets}). Right panel: Group
  velocity, $v^s_g$, of the propagating probe state as a function of
  the probe momentum $k_{pb}$. The black dots are determined from
  simulations, whereas the solid (red) line is the derivative of the
  LP dispersion evaluated at $k_{pb}$, $v^{LP}_{k_{pb}}$. The blue
  dashed line is a guide for eye to indicate where the LP dispersion
  deviates from the quadratic. Adapted from \cite{szymanska10}.}
\label{fig:velog}
\end{figure}

However, we have found conditions under which both signal and
conjugate states get initially strongly amplified by the parametric
scattering from the pump, then decay slowly and, only at later times,
decay exponentially (see Figs.~\ref{fig:bulets}
and~\ref{fig:total_int}) --- we refer to this as the `proper' TOPO
regime. A similar behaviour has been also observed in
experiments~\footnote{\ See, for example, Fig.~3 of
  Ref.~\cite{tosi10}, where the intensity maximum of the extra
  population is reached within $4$~ps after the maximum of the pulsed
  probe, is followed by a slow decay.
}. Now, the spectrum is observed to be linear, $\omega =
\vect{v}_g\cdot \vect{k}$ (see Fig.~\ref{fig:bulets}). A linear
spectrum can be explained by the fact that, in order to have efficient
parametric scattering, probe and conjugate state must have a large
spatial overlap and therefore similar group velocities. Thus signal
and conjugate group velocities need to lock, which is only possible if
the dispersion becomes linear --- a similar result has been also found
in 1D simulations ({\tt see Fabrice Laussy's Chapter}), as well as in
experiments~\cite{amo09}.
The group velocity is defined as the derivative of the energy
dispersion with respect to the momentum. However, we can also measure
it from the probe and the conjugate density variations in time, i.e.,
as $\vect{v}^{pb,c}_g= d \vect{r_m^{pb,c}} /dt$, where
$\vect{r_m^{pb,c}}$ is the maximum of either the probe or conjugate
spatial profile, which we use as a reference.
By analysing the change in time of the spatial profile of the TOPO
probe state $|\psi_{C,X}^{pb}(\vect{r},t)|$, it is possible to
show~\cite{szymanska10} that its group velocity $v_g^{pb}$ is given
exactly by the derivative of the lower polariton (LP) dispersion
evaluated at $\vect{k}_{pb}$, i.e., for zero detuning and low
densities by $v^{LP}_{k_{pb}} \equiv k_{pb}/(2m_C)-k^3/(2m_C \sqrt{k^4
  + 4m_C^2 \Omega_R^2})$ (see Fig.~\ref{fig:velog}).  This behaviour
is consistent with the form of the spectrum shown in
Fig.~\ref{fig:bulets}~\footnote{\ In the regime where the probe
  generates only a weak parametric scattering, aside the strong
  emission from the pump state, the dispersion is simply that of the
  LP, and thus is not surprising that the signal propagates with a
  group velocity given by $v^{LP}_{k_{pb}}$. Remember that here the cw
  pump is below threshold for OPO.}. Further, we have been able to
determine~\cite{szymanska10} that the TOPO linear dispersion is
tangential to the LP branch at $k_{pb}$, thus its slope is given in
this case also by $v^{LP}_{k_{pb}}$.

From the PL spectrum we can also deduce the nature of the wave-packet
propagation. For systems characterised by a linear dispersion, like in
the TOPO regime, one expects a soliton-like behaviour, where probe and
conjugate states propagate without changing neither their shape nor
intensity.
For quadratic dispersion, a Gaussian wave-packet moves at a constant
velocity $v^{LP}_{k_{pb}}= \frac{\vect{k}_{pb}}{2m_C}$ and it
preserves its overall shape in time but its width grows
(FWHM=$(\sigma^2_{pb}+(\frac{t}{2m_C
  \sigma_{pb}})^2)^{1/2}$)~\cite{pitaevskii03}. Note, however, that,
due to the finite polariton lifetime, the total density decays
exponentially, with a rate given by $\frac{\kappa_C+\kappa_X}{2}$ at
zero detuning.
Finally, for non-quadratic dispersion, propagation becomes complex:
The wave-packet gets distorted and there are beatings in the spatial
profiles.
In general, due to the dynamical nature of the TOPO state, the system
evolves between these different scenarios. In particular, only in the
strong amplification regime the spectrum is linear, while it evolves
back to the LP one at longer times. However, for a one-dimensional version
of the equations~\eqref{eq:ppump}, and for uniform, infinitely
extended in space, pumping spots, non-decaying, soliton solutions have
been recently found~\cite{egorov09}. This has been also generalised to
a two-dimensional infinite systems~\cite{egorov10}, where, for some
narrow range of pumping strengths, a soliton-like behaviour has been
predicted for $k_{pb}=0$ and $k_{pb}=k_p$. However, to date, a
non-decaying wave-packet propagation has not been found in
experiments.

Finally we would like to note that, the typical behaviour of probe and
conjugate states (left panel of Fig.~\ref{fig:velog}) is analogous to
the one discussed in four-wave-mixing
experiments~\cite{boyer07,boyer08}: when the probe arrives, and
shortly after that, the conjugate propagates faster then the probe,
before getting locked to it with a small spatial shift of their
maximum intensities. At later times, when the density drops and the
parametric process becomes inefficient, the two wave-packets start
unlocking --- the conjugate slows down with respect to the signal if
$k_c<k_{pb}$ as in Fig.~\ref{fig:velog}, or it moves faster when
$k_c>k_{pb}$.

\subsection{Experiments}
The TOPO regime has been recently studied in experiments in
Refs.~\cite{amo09,ballarini09} (for a review see Ref.~\cite{amo10a}).
As previously described, the additional pulsed probe has been used to
create a travelling, long-living, coherent polaritons signal,
continuously fed by the OPO.
A large increase of the signal lifetime has been observed for a pump
intensity approaching and exceeding the OPO
threshold~\cite{ballarini09}. This observation can be explained in
terms of a critical slowing down of the dynamics following appearance
of a soft Goldstone mode in the spectrum close to threshold. It is
also consistent with the nature of wave-packet propagation in systems
with linear dispersion. This has been used to interpret subsequent
experiments, where the linearisation of dispersion leads to the
suppression of weak scattering and therefore to a polariton motion
without any dissipation~\cite{amo09,amo10a} {\tt cross refer to
  Fabrice Laussy' chapter}.
%
%
Due to the finite size of the excitation spot, the travelling TOPO
signal lives only as long as it reaches the edge of the excitation
spot. However, as discussed in detail in Ref.~\cite{amo10a}, in order
to asses the sustainability in time of the TOPO process, both pump
and probe beams can be chosen so that the probe state forms at
$k_{pb}~0$.  In such a case, the polaritons in the probe state are not
travelling and therefore and it is thus possible to measure the
lifetime of the TOPO wave-packet, which is of the order of a
nanosecond. The decay of the TOPO population in time, as well as the
finite lifetime of TOPO pulses, indicate that the soliton behaviour
predicted in Ref.~\cite{egorov10} is not the explanation of the
current experiments. However, the linearisation of the system's
dispersion due to the parametric process, as well as the appearance of
the Goldstone mode, provide a sufficient explanation of
dissipationless propagation in free space, as well as frictionless
flow against an obstacle, during the part of the dynamics when
parametric processes are strong and the spectrum linear.

\section{Triggered metastable vortices as a diagnostic of the OPO
  superfluid properties}
\label{sec:trigv}
OPO condensates, as well as polariton condensates pumped incoherently,
share with weakly interacting Bose-Einstein condensates at equilibrium
phenomena like the spontaneous breaking of the phase symmetry and the
appearance of a Goldstone mode (see Sec.~\ref{sec:golds}). However,
being intrinsically non-equilibrium, all polaritonic systems need
continuous pumping to balance the fast decay and maintain a steady
state regime. In strong contrast with equilibrium superfluids, the
ground state of which is flow-less, pump and decay lead to currents
that carry polaritons from gain to loss dominated regions. This can
lead to the spontaneous formation of vortices: The presence of
currents in polariton condensates can lead to the spontaneous
appearance of vortices without invoking any superfluid
properties. This is true for incoherently pumped polaritons in
presence of a confining potential~\cite{lagoudakis08,keeling08,
  nardin10}, as well as for polaritons in the OPO regime, with the
difference that here, even in the absence of disorder or a trapping
potential, the system becomes unstable towards the formation of a
quantised vortex state with charge $m=\pm 1$ (see
Sec.~\ref{sec:stabv}). In addition, the hydrodynamic nucleation of
quantised vortices can appear as a consequence of the collisions of a
moving polariton fluid with an obstacle, as will be briefly discussed
in Sec.~\ref{sec:other}.
Therefore, in general, for polaritonic systems, one has to apply some
care when using the appearance of vortices as a diagnostic for the
superfluid properties of such a non-equilibrium system.

In the case of equilibrium superfluids, the rotation of a condensate
is accompanied, above a critical angular
velocity~\cite{pitaevskii03,hess67}, by the creation of quantised
vortices. Here, vortices are stable as far as the system is kept
rotating and become unstable when the imposed rotation is
halted~\cite{rokhsar97}. However, persistent flow can be observed when
a BEC is confined into a toroidal trap and the quantised rotation is
initiated by a pulsed Laguerre-Gauss
beam~\cite{phillips06,phillips07,ramanathan11}. The toroidal trap is
essential to allow the vortex stability, because of the energy cost of
the vortex core to move through the high density region from the
center of the torus where the density is zero.
The very same idea of questioning the persistency of flow in a BEC via
a pulsed Laguerre-Gauss beam as a diagnostic for superfluidity, can be
applied to polaritons~\footnote{\ Note, however, that even if in the
  atomic and polaritonic case the same Laguerre-Gauss laser field is
  used, the mechanism of spinning the BEC atoms is different from the
  one which rotates polaritons.}. As recently proposed for
non-resonantly pumped polariton condensates in Ref.~\cite{wouters10},
this definition of superfluidity as metastable flow is equally
meaningful in non-equilibrium systems as in equilibrium ones. However,
as we will see, the important difference is that, in the OPO regime,
flow persistency is possible even in a simply connected geometry,
i.e., without the need of a toroidal trap which pins the vortex.

A pulsed Laguerre-Gauss (LG) probe beam carrying a vortex of charge
$m$ can be described by:
\begin{equation}
  F_{pb}(\vect{r},t) = f_{pb} |\vect{r}-\vect{r}_{pb}|^{|m|}
  e^{i m \varphi}   e^{-|\vect{r}-\vect{r}_{pb}|^2/(2\sigma^2_{pb})}
  e^{i (\vect{k}_{pb} \cdot \vect{r} - \omega_{pb} t)}
  e^{-(t-t_{pb})^2/(2\sigma^2_{t})} \; ,
\label{eq:vprob}
\end{equation}
with $\{\vect{k}_{pb},\omega_{pb}\}$ can be tuned resonantly to either
the OPO signal or idler. As discussed in the next section, by using a
pulsed LG beam~\eqref{eq:vprob}, vorticity has been shown to persist
not only in absence of the rotating drive, but also longer than the
gain induced by the probe, and therefore to be transferred to the OPO
signal, demonstrating metastability of quantised vortices and
persistence of currents in OPO. Experiments and theory will be
discussed in the next Sec.~\ref{sec:metas}.

\begin{figure}
\begin{center}
\includegraphics[width=0.8\linewidth,angle=0]{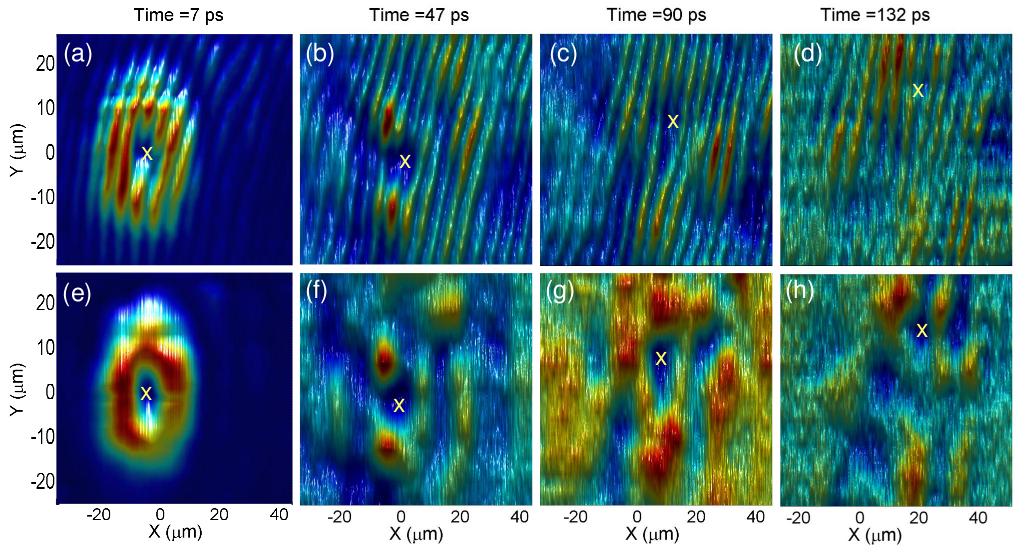}
\end{center}
\caption{Time evolution of the polariton signal following the arrival
  of a LG pulsed beam carrying a vortex of $m=1$. The first raw are the
  interference images obtained by overlapping the vortex with a small
  expanded region of the same image far from the vortex core, where
  the phase is constant, while the second raw are the space profiles
  of the signal. The sequence demonstrates that the vortex remains
  steady as a persisting metastable state for times much longer than
  the extra population created by the probe pulse and eventually gets
  imprinted in the steady state of the OPO signal. This is revealed by
  the strong contrast of the fork in the interference images for as
  long as the core remains within the condensate area. From
  Ref.~\cite{sanvitto10}.}
\label{fig:natph}
\end{figure}
%
\subsection{Theory and experiments}
\label{sec:metas}
%
%
In the case of metastable vortex solutions, the symmetric vortex-less
OPO steady state is dynamically stable, but, because of its superfluid
properties, can support persistent metastable currents injected
externally. From a theoretical point of view, metastable solutions can
be equally induced by either a vortex probe pulse~\eqref{eq:vprob} or
a noise pulse. However, differently from the case of stable vortices,
metastable solutions require a threshold in the perturbation breaking
the system $y \mapsto -y$ symmetry.
For the simulations of Ref.~\cite{sanvitto10}, as we were interested mainly
into the transfer of angular momentum from the probe into the OPO
signal and idler, we have been considering conditions 
where the parametric scattering induced by the probe is too weak to
induce any significant long-lasting amplification, and the gain
introduced by the probe on top of the OPO disappears quite quickly.
We have found conditions where the vortex is transferred from the
probe into the signal~\footnote{\ We checked that $m=\pm 1$ ($m=\mp
  1$) vortex solutions can appear only into the OPO signal (idler). A
  vortex probe pulse of any charge $m$ injected resonantly to the pump
  momentum and energy gets immediately transferred to an $m=\pm 1$
  ($m=\mp 1$) vortex in the signal (idler), leaving the pump
  vortex-less.} (and antivortex in the idler)
immediately~\footnote{\ Later, in Sec.~\ref{sec:topom}, in connection
  to the stability of multiply quantise vortices, we also describe
  vortices in the TOPO regime, where we follow the vortex dynamics not
  of the OPO like here, but of the extra population only.} when the
probe is shined. The transfer is followed by a transient time during
which the imprinted vortex drifts around inside the signal and in
certain cases settles into a metastable solution.
Similarly to what happens to stable vortex solutions, we have found
that the spatial position of the metastable steady state vortices is
close to the position where the OPO signal has the currents pointing
inwards (see second panel of Fig.~\ref{fig:opocl}). The influence of
currents on the formation of vortices is discussed further in
Sec.~\ref{sec:onset}.
Such metastable solutions do not always exist: if the probe is
positioned well inside a wide OPO signal, as the creation of a vortex
is accompanied by the creation of an antivortex (see
Sec.~\ref{sec:onset}), often, the vortex-antivortex pair quickly
recombines; in other cases, during the transient period, the excited
vortex can spiral out of the signal.
Finally note that, as discussed in Sec.~\ref{sec:heali}, the shape and
the size of metastable vortices are independent on the external probe
but are only determined by the parameters of the OPO.

In the experiment of Ref.~\cite{sanvitto10} also shown in
Fig.~\ref{fig:natph}, a vortex is excited by a probe smaller than that
of the signal to allow free motion of the vortex within the
condensate. Vortices are detected, and their evolution in time
followed by a streak camera, in interference images, generated by
making interfere the OPO signal with a constant phase reference beam
in a Michelson interferometer (second row of Fig.~\ref{fig:natph}).
As single shot measurements would give a too low signal to noise
ratio, every picture is the result of an average over many pulsed
experiments taken always for the same OPO conditions.
The probe triggers a TOPO response, creating a strong gain and an
extra decaying population on top of the OPO signal (TOPO). In
experiments, different regimes have been investigated. In particular,
it has been possible to establish that, only under very high pump
power and at specific points in the sample, the vorticity was
transferred from the TOPO into the OPO signal, generating a metastable
vortex solution.  This not only demonstrates that the OPO polariton
condensate can show unperturbed rotation, but also that a vortex can
be another metastable solution of the final steady state,
demonstrating therefore the superfluid behaviour in the
non-equilibrium polariton OPO system. After the vortex is imprinted
into the OPO signal, it has been possible to observe the vortex core
slowly drifting, changing in shape and moving with different
velocities.
Note that, because these are metastable solutions, a minimum probe
power is required for the polaritons to acquire enough angular
momentum to be able to transfer it to the steady state. However, once
the transfer is achieved, the probe power does not change
significantly the duration and depth of the vortex in the steady
state.

\subsection{Onset and dynamics of vortex-antivortex pairs}
\label{sec:onset}
\begin{figure}
\begin{center}
\includegraphics[width=0.7\linewidth,angle=0]{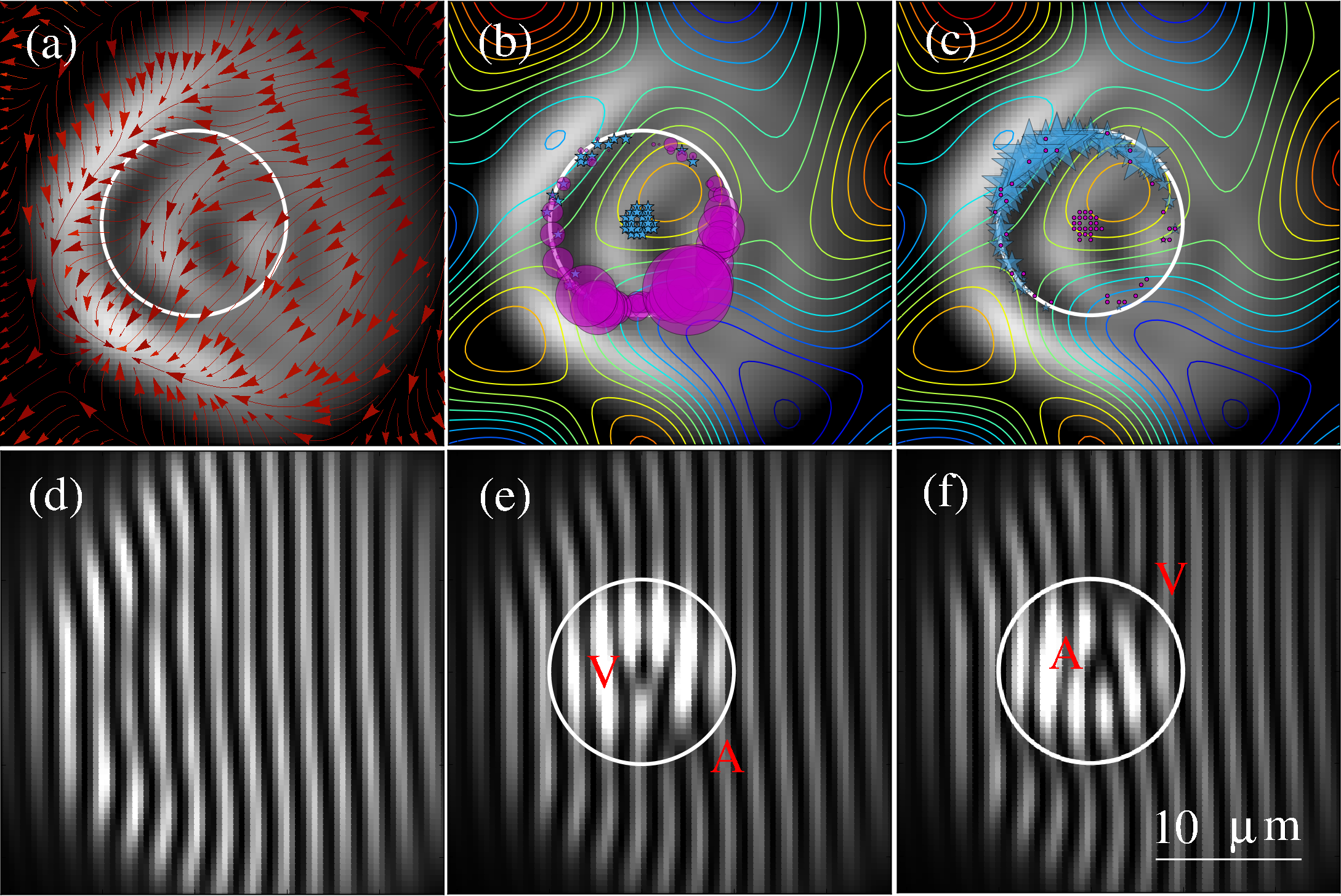}
\end{center}
\caption{Profile and currents of the steady state OPO signal before
  the arrival of the probe (a) and associated interference fringes (d)
  --- parameters for OPO are exactly the same as the ones of the inset
  2 in Fig.~\ref{fig:thres} ($f_p = 1.2f_p^{\text{th}}$).  Location of
  antivortices (dots (b)) and vortices (stars (c)) at the arrival of a
  vortex (stars (b)) or an antivortex (dots (c)) probe, for 1000
  realisations of the random relative phase between pump and probe,
  $\Phi_{rdm}$. The size of dots in (b) (stars in (c)) is proportional
  to the number of times the antivortices (vortices) appear in that
  location. Panel (e) ((f)) shows single shot interference fringes
  relative to the plot in (b) ((c)). Contour-level lines in (b) and
  (c) represent the photonic disorder $V_C(\vect{r})$.  The white
  circle represents the edge of the probe. From Ref.~\cite{tosi11}
  {\tt ask for copyright permission!}.}
\label{fig:prlto}
\end{figure}

There is an aspect that we have been neglecting in the discussion of
the previous section on the occurrence of metastable vortex solutions
in OPO triggered by an external LG probe. If the extension of the
probe carrying a vortex with charge $m=+1$ is smaller than the size of
the vortex-free OPO signal, continuity of the polariton wavefunction
requires that necessarily an antivortex with charge $m=-1$ has to form
at the edge of the probe (see Fig.~\ref{fig:prlto}). Indeed,
`unintended' antivortices have been shown to appear in the signal at
the edge of the imprinting vortex probe and we have explained in
Ref.~\cite{tosi11}, both theoretically and via experiments, the origin
of the deterministic behaviour of the antivortex onset and dynamics,
i.e. where antivortices are more likely to appear in terms of the
currents of the imprinting probe and the ones of the underlying OPO.

%
\subsubsection{Random phase between pump and probe}
As mentioned earlier in Sec.~\ref{sec:metas}, single shot measurements
would give a too low signal to noise ratio, therefore an average is
performed over many pulsed experiments taken always for the same OPO
conditions. What differs at each probe arrival is the random relative
phase $\Phi_{rdm}$ between pump and probe,
\begin{equation}
  F(\vect{r},t) = F_{p}(\vect{r},t) + F_{pb}(\vect{r},t)
  e^{i\Phi_{rdm}} \; ,
\label{eq:randp}
\end{equation}
with $\Phi_{rdm}$ uniformly distributed between $0$ and $2\pi$. We
simulate the dynamics of the vortex-free signal OPO (same conditions
of Fig.~\ref{fig:thres} at $f_p = 1.2f_p^{\text{th}}$) following the
arrival of a vortex probe~\eqref{eq:probe} for 1000 realisations of
$\Phi_{rdm}$ and then average the complex wavefunctions over such
realisations at fixed time and space, $\langle |\psi_{C}^{s}
(\vect{r},t)| e^{i\phi_{C}^{s} (\vect{r},t)}\rangle_{\Phi_{rdm}}$.

The steady state currents of the OPO signal before the arrival of the
probe have a dominant component pointing leftwards and an equilibrium
position where all currents point inwards (bottom left part of the
panel (a) in Fig.~\ref{fig:prlto}). In single shot simulations of
Fig.~\ref{fig:prlto}(d,f) (one realisation of the phase $\Phi_{rdm}$),
we find that if the probe is positioned well inside the OPO signal,
then the imprinting of a vortex $m=+1$ (antivortex $m=-1$) forces the
system to generate, at the same time, an antivortex $m=-1$ (vortex
$m=+1$) at the edge of the probe. This is a consequence of the
continuity of the polariton wavefunctions: If the signal OPO phase is
homogeneous and vortex-free before the arrival of the probe, then
imposing a topological defect, i.e., a branch cut, on the signal phase
at the probe core, requires the branch cut to terminate where the
phase is not imposed by the probe any longer and has to continuously
connect to the freely chosen OPO signal phase, i.e. at the edge of the
probe.
As repeatedly mentioned in this review, OPO parametric scattering
processes constrain the sum of signal and idler phases to the phase of
the laser pump by $2\phi_p=\phi_s+\phi_i$. Thus, at the same positions
where the V-AV pair appears in the signal, an AV-V pair appears in the
idler, so that locally the phase constraint described above is
satisfied. This agrees with the experiments in~\cite{krizhanovskii10},
though there only a single V (AV) in the signal (idler) could be
detected, because the signal size was comparable to the probe one.

Different relative phases $\Phi_{rdm}$ cause the antivortex (vortex)
to appear in different locations around the vortex (antivortex)
probe. However, on 1000 realisations of the random phase uniformly
distributed between $0$ and $2\pi$, we observe that the antivortices
(vortices) are more likely to appear on positions where the current of
the steady state OPO signal before the probe arrival and the probe
current are opposite. For example, for the $m=+1$ ($m=-1$) probe of
Fig.~\ref{fig:prlto}(c) (Fig.~\ref{fig:prlto}(e)), the current
constantly winds anti-clockwise (clockwise), therefore, comparing with
the signal current of Fig.~\ref{fig:prlto}(a), the two are
anti-parallel in the bottom right (top left) region on the probe edge,
region where is very likely that an antivortex (vortex) is
formed. Note also that the onset of antivortices (vortices) privileges
regions where the steady OPO signal has a minimal intensity.
This agrees remarkably well with it has been recently measured
experimentally, in Ref.~\cite{tosi11}.

\subsubsection{Multi-shot averaged dynamics}
\label{sec:aver}
Crucially, via numerical simulations, we elucidate the reason why an
experimental average over many shots allows detecting a vortex by
direct visualisation in density and phase profiles. Recently, it has
been suggested by stochastic simulations~\cite{wouters10} that
vortices in non-resonantly pumped polariton condensates undergo a
random motion which will hinder their direct detection, unless they
are close to be pinned by the stationary disorder potential and thus
follow a deterministic trajectory~\cite{lagoudakis10}.
In the case considered here of a superfluid generated by OPO, we can
instead explain a deterministic dynamics of the V-AV pair in terms of
the OPO steady state currents, which determine a unique
trajectory for the pair, allowing their observation in multi-shot
measurements.

By averaging the 1000 images obtained at the probe arrival, e.g., in
Fig.~\ref{fig:prlto}(c), neither the imprinted vortex nor the
antivortex can be detected: Both phase singularities are washed away
by averaging the differently positioned branch-cuts. However, the
steady state signal currents push the V and AV, initially positioned
in different locations, towards the same equilibrium position where
all currents point inwards. Thus, exactly at the time where the probe
is shined, on average there is no V-AV pair, after $\sim 10$~ps, both
V and AV appear and last $\sim 75$~ps (see Ref.~\cite{tosi11}), till
they eventually annihilate.

It is interesting to note that it has been experimentally
shown~\cite{tosi11} that the onset of vortices in polariton
superfluids does not require a LG imprinting beam, but instead
vortex-antivortex pairs can be also generated when counter-propagating
currents are imposed, similarly to what happens in normal (classical)
fluids. In Ref.~\cite{tosi11} a Gaussian pulsed beam has been shined
either at rest with respect to the OPO signal,
$\vect{k}_{pb}=\vect{k}_s \simeq 0$, or moving $\vect{k}_{pb}\ne
\vect{k}_s$.  While no vortex-antivortex pair appears in the first
case, in the second, a vortex-antivortex pair appears on opposite
sides of the probe edge.

\section{Stability of multiply quantised vortices}
\label{sec:stabi}
%
%
The energy of a vortex is proportional to its quantum of circulation
squared~\cite{pitaevskii03}, $m^2$. Thus, ignoring interactions, a
doubly charged $m=2$ vortex, has higher energy than two single $m=1$
vortices. However, including interactions between vortices, the energy
of an $m=2$ vortex turns to be the same as the energy of two $m=1$
interacting vortices close together. 
The behaviour of doubly quantised vortices has been the subject of
intensive research in the context of ultra-cold atomic gases. In
particular, it has been established that the nature of the splitting
is the dynamical instability. 
Nevertheless, $m=2$ vortices have been
predicted to be stable for specific ranges of density and interaction
strength~\cite{pu99,machida03}, though, so far, they have not been
observed experimentally~\cite{shin04}.
As for single vortices, multiply quantised vortices can be however
stabilised in multiply connected geometry. Indeed, stable pinned
$m=2$ persistent vortices have been recently
observed~\cite{phillips07} by using a toroidal pinning potential
generated by an external optical plug, and demonstrated to split soon
after the plug was removed. In this case, the presence of a plug beam
at the vortex center can pin both $m=1$ and $m=2$ vortex states and
stabilise them against respectively spiralling out of the condensate
for $m=1$ and splitting for $m=2$. In other words, the external trap
mechanically prevents the persistent flow to undergo any movement.

In contrast to equilibrium superfluids, such as atomic gases, both
stable and unstable $m=2$ vortices has been experimentally realised in
polariton OPO superfluids~\cite{sanvitto10}. In this section we
provide a theoretical explaination of the stability and splitting of
doubly charged polariton vortices.
As done previously, vortices in OPO are generated by an external
pulsed probe~\eqref{eq:vprob}. As such, we classify the response of
the system to an $m=2$ LG probe, depending whether the probe generates
a TOPO state (and the vortex is only carried by the extra population
but is not transferred into the OPO signal), as described in
Sec.~\ref{sec:topom}, or instead is transferred in the OPO signal
(Sec.~\ref{sec:opomv}).

\subsection{TOPO regime}
\label{sec:topom}
\begin{figure}
\begin{center}
\includegraphics[width=1.0\linewidth,angle=0]{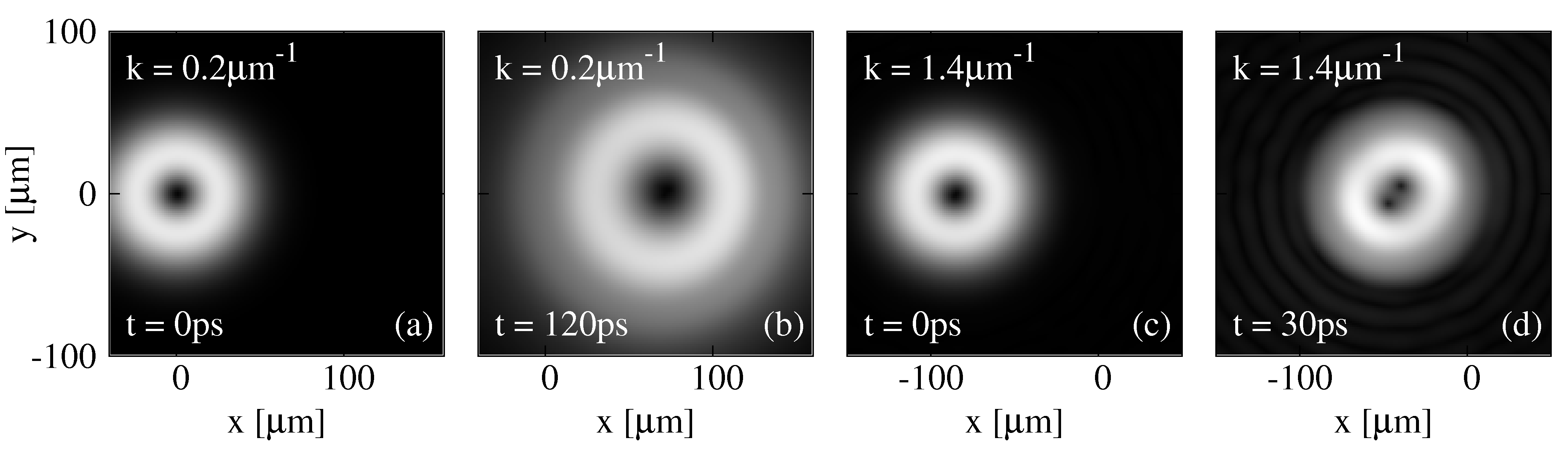}
\end{center}
\caption{Intensity of the TOPO probe profile $|\psi_C^{pb}
  (\vect{r},t)|$ after the arrival (at $t=0$) of an $m=2$ vortex
  pulsed probe~\eqref{eq:vprob} with $\sigma_{pb} \simeq
  87\mu$m. Small $k_{pb}$ is shown in panels (a) and (b), while a
  large $k_{pb}$ in panels (c) and (d). While in the case (a,b) the
  $m=2$ vortex does not split within its lifetime, in (c,d) the vortex
  splits soon after the probe arrives.  The intensity scale in (b) is
  200 times smaller then in (a) -- signal expands as the density drops
  two orders of magnitude (see text). Adapted from
  \cite{szymanska10}.}
\label{fig:Lz2TOPO}
\end{figure}

We first consider the TOPO regime (see Sec.~\ref{sec:topor}), i.e.,
when the vortex propagates inside the triggered probe and conjugate
wave-packets. It has been found~\cite{sanvitto10,szymanska10} that, in
the TOPO regime, $m=2$ vortices are stable within their lifetime when
triggered at small momenta $k_{pb}$ (see Fig.~\ref{fig:Lz2TOPO} panels
(a) and (b)), while they split into two $m=1$ vortices for large
values of $k_{pb}$ (see Fig.~\ref{fig:Lz2TOPO} panels (c) and
(d)). This conclusion was reached both by experimental
observations~\cite{sanvitto10} and theoretical
analysis~\cite{sanvitto10,szymanska10}. The numerical analysis shows
that the crossover from non-splitting to splitting happens for the
probe momenta where the LP dispersion deviates from the quadratic one
(see Fig.~\ref{fig:velog}).
The two different cases are shown in Fig.~\ref{fig:Lz2TOPO}: For
$k_{pb}$ = 0.2$\mu$m$^{-1}$, at short times, the probe propagates
without changing shape and with little change in intensity (not shown
in the Fig.~\ref{fig:Lz2TOPO}), consistent with the linear dispersion
of spectrum characterising this regime. However, at longer times the
density of the triggered probe and conjugate states drops more then
two orders of magnitude, the dispersion changes to the quadratic one
and the wave-packet expands (panel (b)). A uniform expansion of the
wave-packet leads to the decrease of the probe and the conjugate
polariton densities and thus to an increase of the vortex core, but it
does not cause the vortex to split. In contrast, for $k_{pb}$ =
1.4$\mu$m$^{-1}$, where the LP dispersion is not quadratic, the $m=2$
vortex state splits into two $m=1$ vortices shortly after the arrival
of the probe (panel (d)).
This behaviour can be understood by analysing the evolution of the
system's excitation spectrum in time: the dispersion of the
time-dependent TOPO evolves from LP (before the probe arrival) to
linear (at early times after the probe arrival, when the stimulated
scattering is strong), and back to the LP at later times.  For large
$k_{pb}$ the LP dispersion deviates strongly from quadratic (see
Fig.~\ref{fig:velog} right panel).  Wave-packets propagating with
non-quadratic dispersion do not keep their shapes (as discussed in
Sec.~\ref{sec:thtop}), and the simulations show that the distortion
can be very pronounced in particular at later times of the
evolution. The distortion during the early times of the propagation
leads to the mechanical splitting of an $m=2$ vortex, analogous to the
structural instability discussed in
Ref.~\cite{garciaripoll01}. Additionally, as discussed in
Ref.~\cite{sanvitto10}, for small $k_{pb}$, within the quadratic part
of the dispersion, the group velocity of the wave-packet carrying the
vortex equals the velocity of the net super-current (given by
$k_{pb}$) associated with phase variations. This is not however the
case for larger $k_{pb}$, beyond the quadratic part of the
dispersion. In this case, the propagating vortex feels a net current
in its moving reference frame, which may provide additional mechanism
for splitting.
%
%

\subsection{OPO regime}
\label{sec:opomv}
\begin{figure}
\begin{center}
\includegraphics[width=1.0\linewidth,angle=0]{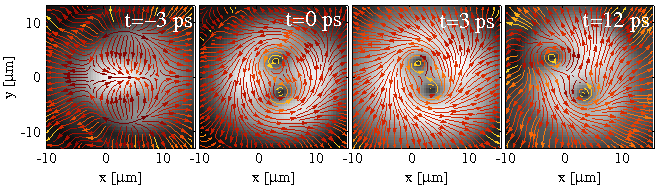}
\end{center}
\caption{Filtered signal profile and currents above threshold for OPO
  ($f_p=1.12 f_p^{\text{(th)}}$) for a top-hat pump with FWHM
  $\sigma_p=35$~$\mu$m at $t=-3$~ps (first panel) before the arrival
  of an $m=2$ probe. The doubly quantised vortex gets transferred from
  the probe into the OPO signal and splits into two $m=1$ vortices
  even before the probe reaches it's maximum intensity at $t=0$~ps
  (second panel). In this simulation the vortices coexist for sometime
  (roughly $15$~ps), then one gets expelled from the signal (fourth
  panel). Adapted from~\cite{szymanska10}.}
\label{fig:Lz2OPO}
\end{figure}

In contrast to the TOPO regime described above, it has been shown both
experimentally and theoretically~\cite{sanvitto10,szymanska10} that
$m=2$ vortices that do get imprinted into the steady-state OPO signal
are never stable and splits into two $m=1$ vortices almost
immediately, even before the probe reaches its maximum intensity (see
Fig.~\ref{fig:Lz2OPO}). By analysing the system's dispersion in
different regimes, as well as the dynamics of currents visible in the
simulations, we have been able to identify several causes for the
splitting:
Before the arrival of the probe, the steady-state OPO dispersion is
flat around the pump, signal, and idler. However, the triggering probe
favours the signal and conjugate to lock and propagate with the same
velocity $v^{LP}_{k_{pb}}$. This behaviour corresponds to a linear
dispersion. Further, once the vortex gets imprinted into the
stationary OPO signal and idler, the system's dispersion changes back
to be flat.  The evolution of the dispersion between flat, linear and
again flat leads to a complicated dynamics of both signal and idler
(the \emph{transient period} described in~\cite{marchetti10}), causing
the structural instability and splitting of the $m=2$ vortex during
the transient time. Another reason for the structural instability and
splitting are the non-uniform currents (see Fig.~\ref{fig:Lz2OPO})
present in the OPO signal caused by the interplay between spatial
inhomogeneity, pump and decay, which the OPO vortex experiences in its
reference frame.

\section{Vortices in other polariton fluids}
\label{sec:other}
We do not pretend to give an exhaustive review of the broad field
which studies vortices in polariton fluids in this chapter, where we
have mostly focussed on the occurrence of vortices in polariton OPO
superfluids. However, we would like at least to briefly mention what
happens for polariton fluids other than OPO.

\paragraph{Spontaneous vortices in trapped incoherently pumped polaritons}
For incoherently pumped polaritons, the presence of a harmonic
trapping potential, can make the non-rotating solution unstable to the
spontaneous formation of a vortex lattice ~\cite{keeling08} (this work
has been generalised to include the effects of polarisation in
Ref.~\cite{borgh10}). In experiments, spontaneous vortices in
incoherently pumped systems, have been observed in
Refs.~\cite{lagoudakis08,lagoudakis11} and their existence explained
in terms of pinning by the disorder present in the CdTe sample.
Adding the polarisation degrees of freedom, can give rise to the
appearance of half-vortices~\cite{Rubo07,lagoudakis09a}.
Polariton vortices have been also observed in cavity
mesas~\cite{nardin10}. {\tt cross refer to Benoit Deveaud's chapter?}

Vortex-antivortex pairs have been observed in the non-resonantly
pumped experiments of Ref.~\cite{roumpos10}, where the mechanism of
V-AV generation is explained in terms of density fluctuations
originating from the cw multi-mode pumping laser, while for a
single-mode laser no V-AV pairs have been observed. In this sense,
their motivation and interpretation is in terms of the BKT
transition. The pair dynamics in the condensate has been studied in
Ref.~\cite{fraser09}. {\tt cross refer to Yoshisha Yamamoto's chapter?}

Finally, as mentioned previously in Secs.\ref{sec:trigv}
and~\ref{sec:aver}, generation and detection of metastable vortices
have been also recently discussed for polariton condensates generated
by toroidal non-resonantly pumping in Ref.~\cite{wouters10}, where
vortices have been seeded with an external LG probe.
Interestingly, very recently in Ref.~\cite{mann11}, it has been
observed an all-optical spontaneous pattern formation in a polariton
condensate non-resonantly pumped with a ring geometry.

\paragraph{Resonantly pumped-only polaritons}
In Ref.~\cite{nardin11,sanvitto11}, the hydrodynamic nucleation of
V-AV pairs is studied by making collide the polariton fluid with a
large defect. In particular, polaritons are resonantly (coherently)
injected with a pulsed laser beam, creating a population in the pump
state only. 
The focus and interest of these studies are the possibility of
exploring quantum turbulence, the appearance of dissipation and drag
above a critical velocity because of the nucleation of vortices in the
wake of the obstacle.


\begin{acknowledgement}
  We would like to acknowledge fruitful collaborations with
  A. Berceanu, E. Cancellieri, D. Sanvitto, C. Tejedor, and
  D. M. Whittaker on some of the topics discussed in this review, as
  well as the collaboration with the experimental group at UAM in
  Madrid (C. Ant\'on, M. Baudisch, G. Tosi, L. Vi\~na). We are
  particularly grateful to I. Carusotto and J. Keeling for stimulating
  discussions and for the critical reading of this manuscript. We also
  would like to thank C. Creatore, B. Deveaud-Pl\'edran, and
  Y. Yamamoto for useful suggestions and points of
  discussion. F.M.M. acknowledges the financial support from the
  programs Ram\'on y Cajal and POLATOM (ESF). This work has been also
  supported by the Spanish MEC (MAT2008-01555, QOIT-CSD2006-00019),
  CAM (S-2009/ESP-1503) and FP7 ITN "Clermont4" (235114).
\end{acknowledgement}

\newcommand\textdot{\.}

\end{document}